\documentstyle[preprint,aps,epsf,amsfonts]{revtex}
\tighten
\begin{document}
\draft
%@twocolumnfalse\endcsname

% ****************************** TITLE *******************************

\title{Late-time decay of gravitational and electromagnetic
 perturbations along the event horizon}
\author{Leor Barack \and Amos Ori}
\address {Department of Physics,
          Technion---Israel Institute of Technology, Haifa, 32000, Israel}
\date{\today}
\maketitle

%****************************  ABSTRACT *********************************

\begin{abstract}
We study analytically, via the Newman-Penrose formalism,
the late-time decay of linear electromagnetic and gravitational
perturbations along the event horizon (EH) of black holes.
We first analyze in detail the case of a Schwarzschild black hole.
Using a straightforward local analysis near the EH, we show that,
generically, the ``ingoing'' ($s>0$) component
of the perturbing field dies off along the EH more rapidly than
its ``outgoing'' ($s<0$) counterpart.
Thus, while along $r=const>2M$ lines both components
of the perturbation admit the well-known $t^{-2l-3}$
decay rate, one finds that along the EH the $s<0$ component
dies off in advanced-time $v$ as $v^{-2l-3}$,
whereas the $s>0$ component dies off as $v^{-2l-4}$.
We then describe the extension of this analysis to a Kerr black hole. We
conclude that for axially symmetric modes the situation is analogous
to the Schwarzschild case. However, for non-axially symmetric modes
both $s>0$ and $s<0$ fields decay at the same rate (unlike in the
Schwarzschild case).
\end{abstract}


%\vspace{6ex}
%]
% ************************************************************************

\section{introduction}

When a gravitational collapse results in the formation of
a black hole (BH), the gravitational field outside the the event
horizon (EH) relaxes at late time to the stationary Kerr-Newman geometry.
Also, when the (pure) Kerr-Newman field external to a BH is perturbed
by gravitational or electromagnetic waves, the perturbing field
die off at late time everywhere outside the BH, and along
its EH.
In both scenarios, it is implied by the ``{\em no hair}'' principle
that when the BH geometry settles down into its stationary state,
all characteristics of the initial state (or initial perturbation)
must somehow be lost, except for the conserved quntities associated
with it: its total mass, electric charge, and angular momentum.
(For a detailed review of the ``no hair'' theorems by Hawking,
Israel, Carter, and Robinson, see \cite{Carter73}.)

Remarkable as the ``no-hair'' principle is, it still gives
no information about the mechanism through which
this ``compulsory'' relaxation process occurs. For example,
it tells us nothing about the rate of the decay process.
Clearly, such a detailed description of the late time decay
is important not only for gaining more insight into the ``no hair''
principle, but, more practically, by virtue of the
recent prospects of detecting gravitational radiation from
astrophysical black hole systems.
Also, the characteristics of the decay along
the event horizon has an impact on the nature of the
singularity along the inner horizon of charged \cite{InnerCharged}
and rotating \cite{InnerRotating} black holes.

A detailed description of the late time decay outside
a Schwarzschild black holes was first given by Price
(for scalar and metric perturbations\cite{PriceI}, and
for all integer-spin fields in the Newman-Penrose
formalism \cite{PriceII}). Price found that any radiative multipole
mode $l,m$ of an initially-compact linear perturbation dies off at
late time as $t^{-2l-3}$ (where $t$ is the Schwarzschild time coordinate).
If a static multipole mode existed prior to the formation of the BH,
then this mode will decay as $t^{-2l-2}$.
Price found these power law decay tails to be the same for all kinds of
perturbations, whether scalar, electromagnetic or gravitational (and in
this respect, the scalar field model proved as a useful
toy-model for the more realistic fields).

Price's results were later confirmed using several different
approaches, both analytic and numerical
\cite{Leaver86,Gundlach94I,Winicour94,Andersson97,Barack99I,Barack99II},
and where generalized to other spherically symmetric spacetimes
\cite{Gundlach94I,Bicak72,Ching95,Burko97,Brady97,Contamination,Piran}.
The validity of the perturbative (linear) approach
was supported by numerical analyses of the fully nonlinear
dynamics \cite{Gundlach94II,Burko97}, indicating virtually the
same power law indices for the late-time decay.

Recently, several authors addressed the issue of the late-time
decay of fields outside rotating black holes.
First, a numerical simulation of the evolution of linear scalar
\cite{Krivan96} and gravitational \cite{Krivan97} waves
on the background of a Kerr black hole was carried out
by Krivan {\em et al}. Later, an analytic treatment of this problem
(in the time domain) was presented by Barack and Ori
\cite{Proc,Letter,Kerr} (following a preliminary analysis by Ori
\cite{Amos}). Then, a study of the late time decay in
Kerr using a frequency-domain approach has been carried
out by Hod, both for a scalar field \cite{HodI} and for nonzero-spin
Newman-Penrose fields \cite{HodII} (following preliminary
considerations by Andersson \cite{Andersson97}).

The above analyses all indicate that power law tails characterize the
decay in the Kerr background as well.
In this case, however, the lack of spherical symmetry causes coupling
between various multipoles. As a result of this coupling, the
power-law indices of specific spherical-harmonics multipoles
are found to be different, in general, from
the ones obtained in spherically symmetric black holes.
Another phenomenon caused by rotation (first observed in
\cite{Amos}) is the oscillatory nature of the late time tails
along the null generators of the EH of the Kerr BH for
nonaxially-symmetric perturbation modes.
(See \cite{Letter,Kerr} for details.)

As we just mentioned, power law tails are observed not only
at timelike infinity, but also at future null infinity and
along the (future) event horizon.
Several authors have analyzed the late time behavior of a scalar
field along the EH of a Schwarzschild BH
\cite{Gundlach94I,Barack99II} and a Kerr BH \cite{Letter,HodI}.
In both cases, the power law indices of the late-time decay
along the EH were found to be the same as along any fixed-$r$
world-line outside the BH (apart from the above mentioned
oscillations along the EH in the Kerr case).
Thus, in the Schwarzschild case, an $l,m$ scalar perturbation mode is
found to decay along the EH as $v^{-2l-3}$ (or
$v^{-2l-2}$ for an initially static mode).
(Here, $v$ is an advanced-time coordinate, which we define in
the sequel.)

Quite surprisingly, a careful and thorough study of the
behavior of realistic physical fields (electromagnetic and gravitational)
along the EH has not been carried out so far (to the best
of our knowledge), even in the Schwarzschild case \cite{Hod3}.
One would expect the scalar-field model to provide, again, a reliable
picture of the actual behavior of realistic physical fields;
however, a careful analysis of the behavior of such realistic fields
at the EH reveals several interesting new features, uncovered by the
scalar-field case.
These features arise already in the Schwarzschild case, and thus
we find it instructive to study and explain this simpler case
first.
Accordingly, in this paper we first explore in detail the behavior of
electromagnetic and gravitational perturbations at the EH of a
Schwarzschild BH.
Then we describe the extension of this analysis to the Kerr case, and
derive the power-law indices at the EH. Full detail of the analysis of
the Kerr case will be given in a forthcoming paper \cite{Kerr}
(as part of a comprehensive analysis of the late time decay of
perturbations in the Kerr spacetime).

We shall apply a linear perturbation analysis, based on the
Newman-Penrose formalism. In this framework, a single master
equation governs the (gauge-invariant) radiative parts of the
linear perturbations of both the Maxwell tensor and the Weyl tensor.
For both fields, our analysis reveals that the ``ingoing''
($s>0$) part of the perturbing field dies off at late time along
the EH of the Schwarzschild BH {\em faster} than its ``outgoing''
($s<0$) counterpart: Whereas the $s<0$ fields admit the usual
$v^{-2l-3}$ law, the $s>0$ fields decay at the EH like $v^{-2l-4}$.
In the Kerr case, the above difference in the behavior of the $s>0$
and $s<0$ fields occurs
only for axially symmetric ($m=0$) modes; for non-axially symmetric
modes, one finds the same decay rates for both $s>0$ and $s<0$.
These results are summarized in
Eqs.\ (\ref{eq118m}), (\ref{eq118a}), and (\ref{eq118b}) below,
in the concluding section.
We also comment there about the significance of our results to the
study of the interior of spinning black holes, and discuss the relation
of our analysis to previous works \cite{Hod3}.

An important role in our analysis is played by the static solutions
of the field equation. These turn out to show a peculiarity:
As in the scalar field case, there is a static solution regular
at the horizon, and a second, independent, solution which is
irregular there.
However, for $s>0$ fields, regularity of a static solution cannot be
judged merely from its leading-order behavior at the EH.
Rather, the distinction between the regular and irregular
solutions involves the identification of a certain, sub-dominant,
logarithmic term in the latter.
Another peculiarity has to do with the relation between static
solutions and monochromatic solutions (i.e. modes of a single Fourier
frequency $\omega$).
For $s>0$ fields, unlike the scalar field (and unlike the $s<0$ case), the
EH-regular static solution cannot be approached from an EH-regular
monochromatic solution by naively taking the limit
$\omega\rightarrow 0$. One finds that for $s>0$ this naive limit leads
to a static solution {\em irregular} at the EH.
We study these unexpected features, and then qualitatively explain
them using a simple (scalar-field based) toy model.

The paper is arranged as follows:
In Sec.\ \ref{preliminaries} we give some
definitions and notations, and briefly review the Newman-Penrose
formalism for perturbations of the Schwarzschild geometry.
In Sec.\ \ref{LTE} we introduce the {\em late time expansion},
to be employed in our analysis.
The static solutions to the field equation, central to our
analysis, are obtained in Sec.\ \ref{static}, followed
(in Sec.\ \ref{regularity}) by a formulation of regularity
criteria for physical fields at the EH.
This puts us in position to analyze (in Sec.\ \ref{behavior})
the late time behavior of physical fields along the EH.
This analysis yields the power index for both $s<0$ and $s>0$
fields.
Another perspective on the subject is obtained  in Sec.\
\ref{monochromatic}, where we consider the behavior of momochromatic
modes.
In Sec.\ \ref{toy} we then introduce a simple toy-model, which
yields further insight into our results.
The extension of our analysis to the case
of a Kerr BH is described in Sec.\ \ref{Kerr}.
In the concluding section (sec.\ \ref{conclusions}) we summarize
the results and discuss their physical significance and their
relation to other works.

\section{Definitions and Notations}
\label{preliminaries}

The line element in the Schwarzschild spacetime reads, in the standard
Schwarzschild coordinates $t,r,\theta,\varphi$,
%~~~~~~~~~~~~~~~~~~~~~~~~~~~~~~~~~~~~~~~~~~~~~~~~~~~~~~~~~~~~~~~~~~~~~~
\begin{equation} \label{Metric}
ds^2=-(\Delta/r^2)\,dt^2+(r^2/\Delta)\,dr^2 +r^2 (d\theta^2
    + \sin^2\theta\, d\varphi^2),
\end{equation}
%~~~~~~~~~~~~~~~~~~~~~~~~~~~~~~~~~~~~~~~~~~~~~~~~~~~~~~~~~~~~~~~~~~~~~~
where $M$ is the mass of the BH, and
%~~~~~~~~~~~~~~~~~~~~~~~~~~~~~~~~~~~~~~~~~~~~~~~~~~~~~~~~~~~~~~~~~~~~~~
\begin{equation} \label{Delta}
\Delta(r)\equiv r^2-2Mr
\end{equation}
%~~~~~~~~~~~~~~~~~~~~~~~~~~~~~~~~~~~~~~~~~~~~~~~~~~~~~~~~~~~~~~~~~~~~~~
is a function which vanishes at the EH, $r=2M$.
Here, and throughout this paper, we use relativistic units, $c=G=1$.

As this paper concerns with the behavior near the event horizon
(EH), we shall find it convenient in the sequel
to introduce a new (dimensionless) radial coordinate,
%~~~~~~~~~~~~~~~~~~~~~~~~~~~~~~~~~~~~~~~~~~~~~~~~~~~~~~~~~~~~~~~~~~~~~~
\begin{equation} \label{z}
z\equiv \frac{r-2M}{2M}
\end{equation}
%~~~~~~~~~~~~~~~~~~~~~~~~~~~~~~~~~~~~~~~~~~~~~~~~~~~~~~~~~~~~~~~~~~~~~~
which vanishes at the horizon.

We shall also need the EH-regular (Kruskal) null
coordinates
%~~~~~~~~~~~~~~~~~~~~~~~~~~~~~~~~~~~~~~~~~~~~~~~~~~~~~~~~~~~~~~~~~~~~~~
\begin{equation} \label{Kruskal}
    V\equiv e^{v/(4M)},\;\;\;\;    U\equiv -e^{-u/(4M)},
\end{equation}
%~~~~~~~~~~~~~~~~~~~~~~~~~~~~~~~~~~~~~~~~~~~~~~~~~~~~~~~~~~~~~~~~~~~~~~
where $v\equiv t+r_*$ and $u\equiv t-r_*$ are the Eddington-Finkelstein
null coordinates, with
%~~~~~~~~~~~~~~~~~~~~~~~~~~~~~~~~~~~~~~~~~~~~~~~~~~~~~~~~~~~~~~~~~~~~~~
\begin{equation} \label{r*}
r_*\equiv r+2M\ln z.
\end{equation}
%~~~~~~~~~~~~~~~~~~~~~~~~~~~~~~~~~~~~~~~~~~~~~~~~~~~~~~~~~~~~~~~~~~~~~~

To discuss perturbations of the Schwarzschild BH via the
Newman-Penrose formalism, we introduce
the tetrad basis of null vectors
$\left(l^\mu,n^\mu,m^\mu,m^{*\mu}\right)$, defined in the
($t,r,\theta,\varphi$) coordinate system by \cite{PriceII,Bardeen73}
%~~~~~~~~~~~~~~~~~~~~~~~~~~~~~~~~~~~~~~~~~~~~~~~~~~~~~~~~~~~~~~~~~~~~~~
\begin{eqnarray} \label{tetrad}
l^\mu&=&\left[r^2/\Delta,1,0,0\right] \nonumber\\
n^\mu&=&\left[1,-\Delta/r^2,0,0\right]/2 \nonumber\\
m^\mu&=& \left[0,0,1,i/\sin\theta\right]/(2^{1/2}r).
\end{eqnarray}
%~~~~~~~~~~~~~~~~~~~~~~~~~~~~~~~~~~~~~~~~~~~~~~~~~~~~~~~~~~~~~~~~~~~~~~
(The components of the forth tetrad leg, $m^{*\mu}$, are obtained
from the components of $m^{\mu}$ by complex conjugation.)

In the framework of the Newman--Penrose formalism
\cite{NP} the gravitational field in vacuum is completely described by
five complex scalars, $\Psi_0\ldots\Psi_4$, constructed
from the Weyl tensor by projecting it on the above tetrad basis.
Likewise, the electromagnetic field is completely described
by the three complex scalars $\varphi_0,\varphi_1,\varphi_2$,
constructed by similarly projecting the Maxwell tensor.
In particular,
%~~~~~~~~~~~~~~~~~~~~~~~~~~~~~~~~~~~~~~~~~~~~~~~~~~~~~~~~~~~~~~~~~~~~~~
\begin{eqnarray} \label{Psi}
\Psi_0\equiv -C_{\alpha\beta\gamma\delta}
           l^\alpha m^\beta l^\gamma m^\delta,\,\,\,\,\,
\Psi_4\equiv -C_{\alpha\beta\gamma\delta}
           n^\alpha m^{*\beta} n^\gamma m^{*\delta}
\end{eqnarray}
%~~~~~~~~~~~~~~~~~~~~~~~~~~~~~~~~~~~~~~~~~~~~~~~~~~~~~~~~~~~~~~~~~~~~~~
represent the ingoing and outgoing radiative parts, respectively,
of the Weyl tensor, and
%~~~~~~~~~~~~~~~~~~~~~~~~~~~~~~~~~~~~~~~~~~~~~~~~~~~~~~~~~~~~~~~~~~~~~~
\begin{eqnarray} \label{varphi}
\varphi_0\equiv F_{\mu\nu} l^\mu m^\nu,\,\,\,\,\,
\varphi_2\equiv F_{\mu\nu} m^{*\mu}n^\nu
\end{eqnarray}
%~~~~~~~~~~~~~~~~~~~~~~~~~~~~~~~~~~~~~~~~~~~~~~~~~~~~~~~~~~~~~~~~~~~~~~
represent the ingoing and outgoing radiative parts
of the electromagnetic field.

In the Schwarzschild (unperturbed) background all Weyl scalars
but $\Psi_2$ vanish (as directly implied by the Goldberg-Sachs
theorem, in view of the Schwarzschild spacetime being of Petrov type D;
see Sec.\ 9b,c in \cite{Chandra83}).
In the framework of a linear perturbation analysis,
the symbols $\Psi_0,\Psi_1,\delta\Psi_2,\Psi_3,\Psi_4$ and
$\varphi_0,\delta\varphi_1,\varphi_2$ are thus used to represent
first-order perturbations of the corresponding fields
(with $\delta\Psi_2\equiv \Psi_2-\Psi_2^{\rm{background}}$, etc.).
One can show (see Sec.\ 29b in \cite{Chandra83}) that
$\Psi_0$ and $\Psi_4$, and also $\varphi_0$ and $\varphi_2$,
are {\em invariant} under gauge transformations (namely, under
infinitesimal rotations of the null basis and infinitesimal
coordinate transformations).
The scalars $\Psi_1$ and $\Psi_3$ are not gauge invariant, and
may be nullified by a suitable rotation of the null frame.
The entities $\delta\Psi_2$ and $\delta\varphi_1$ represent perturbations
of the ``Coulomb-like'', non-radiative, part of the fields
(in fact, one can also nullify $\delta\Psi_2$ by a suitable infinitesimal
coordinate transformation.)
It is therefore only the scalars defined in Eqs.\ (\ref{Psi}) and
(\ref{varphi}) which carry significant information about the radiative
part of the fields.
(Note, however, that gauge invariance of the radiative fields
is guaranteed only within the framework of linear perturbation
theory.)

There is a single master equation governing linear perturbations of both
the gravitational and the electromagnetic radiative fields defined in
Eqs.\ (\ref{Psi},\ref{varphi})\cite{Master}.
In vacuum, this master perturbation equation reads
%~~~~~~~~~~~~~~~~~~~~~~~~~~~~~~~~~~~~~~~~~~~~~~~~~~~~~~~~~~~~~~~~~~~~~~
\begin{eqnarray} \label{Master}
r^4\Delta^{-1}\Psi^{s}_{,tt} -
\Delta^{-s}(\Delta^{s+1} \Psi^{s}_{, r})_{,r}
-\frac{1}{\sin\theta}(\Psi^{s}_{,\theta}\sin\theta)_{,\theta}
-\frac{1}{\sin^{2}\theta}\Psi^{s}_{,\varphi\varphi}
     -\frac{2is\cos\theta}{\sin^2\theta}\Psi^{s}_{,\varphi} \nonumber\\
 -2s[Mr^2/\Delta-r]\Psi^{s}_{,t}
 +(s^2\cot^2\theta -s)\Psi^{s}=0,
\end{eqnarray}
%~~~~~~~~~~~~~~~~~~~~~~~~~~~~~~~~~~~~~~~~~~~~~~~~~~~~~~~~~~~~~~~~~~~~~~
where $\Psi^s(t,r,\theta,\varphi)$ represents the various radiative fields
according to the following substitutions:
%~~~~~~~~~~~~~~~~~~~~~~~~~~~~~~~~~~~~~~~~~~~~~~~~~~~~~~~~~~~~~~~~~~~~~~
\begin{eqnarray} \label{fields}
\varphi_0 &=& \Psi^{s=+1}                     \nonumber\\
\varphi_2 &=& r^{-2}\Psi^{s=-1}  \nonumber\\
\Psi_{0}  &=& \Psi^{s=+2}                     \nonumber\\
\Psi_{4}  &=& r^{-4}\Psi^{s=-2}.
\end{eqnarray}
%~~~~~~~~~~~~~~~~~~~~~~~~~~~~~~~~~~~~~~~~~~~~~~~~~~~~~~~~~~~~~~~~~~~~~~

In Eq.\ (\ref{Master}), the angular dependence of $\Psi^s$ is
separable through a decomposition in
{\em spin-weighted spherical harmonics}\cite{Goldberg67},
%~~~~~~~~~~~~~~~~~~~~~~~~~~~~~~~~~~~~~~~~~~~~~~~~~~~~~~~~~~~~~~~~~~~~~~
\begin{equation} \label{eq1}
\Psi^s(r,t,\theta,\varphi)=\sum_{l=|s|}^{\infty}\sum_{m=-l}^{l}\,
       \psi^{slm}(r,t)\, Y^{slm}(\theta,\varphi).
\end{equation}
%~~~~~~~~~~~~~~~~~~~~~~~~~~~~~~~~~~~~~~~~~~~~~~~~~~~~~~~~~~~~~~~~~~~~~~
The time-radial functions $\psi ^{slm}(r,t)$ then satisfy the field
equation
%~~~~~~~~~~~~~~~~~~~~~~~~~~~~~~~~~~~~~~~~~~~~~~~~~~~~~~~~~~~~~~~~~~~~~~
\begin{equation} \label{eq2}
r^4 \psi^{slm}_{,tt} - \Delta^{-s+1}(\Delta^{s+1}\psi^{slm}_{,r})_{,r}
+2sr^2(r-3M) \psi^{slm}_{,t} + (l-s)(l+s+1)\Delta\psi^{slm}=0.
\end{equation}
%~~~~~~~~~~~~~~~~~~~~~~~~~~~~~~~~~~~~~~~~~~~~~~~~~~~~~~~~~~~~~~~~~~~~~~

\section{The late-time expansion}
\label{LTE}

In order to analyze the power-law decay of perturbations at late
time, we decompose the field in the form
%~~~~~~~~~~~~~~~~~~~~~~~~~~~~~~~~~~~~~~~~~~~~~~~~~~~~~~~~~~~~~~~~~~~~~~
\begin{equation} \label{eq4}
\psi ^{slm}(r,t)=\sum\limits_{k=0}^\infty {F_k^{slm}(r)}
\,v^{-k_0-k},
\end{equation}
%~~~~~~~~~~~~~~~~~~~~~~~~~~~~~~~~~~~~~~~~~~~~~~~~~~~~~~~~~~~~~~~~~~~~~~
to which we refer as the {\em late-time expansion}\cite{LTE}.
Substitution in Eq.\ (\ref{eq2}) yields an ordinary equation for
each function $F_{k}^{slm}$:
%~~~~~~~~~~~~~~~~~~~~~~~~~~~~~~~~~~~~~~~~~~~~~~~~~~~~~~~~~~~~~~~~~~~~~~
\begin{equation} \label{eq5}
D^{sl}(F_k^{slm})=S_k^{slm},
\end{equation}
%~~~~~~~~~~~~~~~~~~~~~~~~~~~~~~~~~~~~~~~~~~~~~~~~~~~~~~~~~~~~~~~~~~~~~~
where $D^{sl}$ is a differential operator defined by
%~~~~~~~~~~~~~~~~~~~~~~~~~~~~~~~~~~~~~~~~~~~~~~~~~~~~~~~~~~~~~~~~~~~~~~
\begin{equation} \label{eq6}
D^{sl}\equiv
\Delta d^2/dr^2+2(s+1)(r-M)d/dr-(l-s)(l+s+1),
\end{equation}
%~~~~~~~~~~~~~~~~~~~~~~~~~~~~~~~~~~~~~~~~~~~~~~~~~~~~~~~~~~~~~~~~~~~~~~
and the source term $S_k^{slm}$ is given by
%~~~~~~~~~~~~~~~~~~~~~~~~~~~~~~~~~~~~~~~~~~~~~~~~~~~~~~~~~~~~~~~~~~~~~~
\begin{equation} \label{eq7}
S_k^{slm}\equiv
2(k_0+k-1)r\left[\frac{d(rF_{k-1}^{slm})}{dr} +
2srM\Delta^{-1}F_{k-1}^{slm}\right].
\end{equation}
%~~~~~~~~~~~~~~~~~~~~~~~~~~~~~~~~~~~~~~~~~~~~~~~~~~~~~~~~~~~~~~~~~~~~~~
(We take $F_{k<0}^{slm}\equiv 0$.)

The dominant late-time decay at world-lines of fixed $r$ is
described by the term $k=0$ in Eq.\ (\ref{eq4}). To the leading
order in $1/v$, we have
%~~~~~~~~~~~~~~~~~~~~~~~~~~~~~~~~~~~~~~~~~~~~~~~~~~~~~~~~~~~~~~~~~~~~~~
\begin{equation} \label{eq8}
\psi^{slm}(r,t)\cong F_{k=0}^{slm}(r)\,v^{-k_0}.
\end{equation}
%~~~~~~~~~~~~~~~~~~~~~~~~~~~~~~~~~~~~~~~~~~~~~~~~~~~~~~~~~~~~~~~~~~~~~~
Substituting $v=t+r_*$, we also find, to the leading order in
$1/t$,
%~~~~~~~~~~~~~~~~~~~~~~~~~~~~~~~~~~~~~~~~~~~~~~~~~~~~~~~~~~~~~~~~~~~~~~
\begin{equation} \label{eq9}
\psi^{slm}(r,t)\cong F_{k=0}^{slm}(r)\,t^{-k_0},
\end{equation}
%~~~~~~~~~~~~~~~~~~~~~~~~~~~~~~~~~~~~~~~~~~~~~~~~~~~~~~~~~~~~~~~~~~~~~~
which, using the well-known result by Price \cite{PriceII},
implies $k_0=2l+3$ (or $k_0=2l+2$ if a static mode $l$ is initially
present).\footnote{
For brevity, we hereafter consider modes without initial static
multipoles.}

Since $S_k^{slm}$ vanishes for $k=0$, the term $F_{k=0}^{slm}(r)$
satisfies the homogeneous differential equation
%~~~~~~~~~~~~~~~~~~~~~~~~~~~~~~~~~~~~~~~~~~~~~~~~~~~~~~~~~~~~~~~~~~~~~~
\begin{equation} \label{eq10}
D^{sl}(F_0^{slm})=0.
\end{equation}
%~~~~~~~~~~~~~~~~~~~~~~~~~~~~~~~~~~~~~~~~~~~~~~~~~~~~~~~~~~~~~~~~~~~~~~
This is just the field equation of a {\em static} mode $l,m$.
Thus, $F_{k=0}^{slm}(r)$ must be a static solution of
the field equation. In the next section we shall study the
static solutions for $\Psi^s$, focusing attention on their
asymptotic behavior at the EH.

\section{Static solutions} \label{static}

Since Eq.\ (\ref{eq10}) is a second-order differential equation,
its general solution is spanned by two basis solutions. We
shall primarily be interested in the asymptotic behavior of
these solutions near the EH.
The leading-order asymptotic behavior can be easily obtained from the
asymptotic form of Eq.\ (\ref{eq10}) near the EH: One finds that
for both $s>0$ and $s<0$, the two asymptotic solutions behave there
like
%~~~~~~~~~~~~~~~~~~~~~~~~~~~~~~~~~~~~~~~~~~~~~~~~~~~~~~~~~~~~~~~~~~~~~~
\begin{equation} \label{eq10a}
\psi^{slm}\cong \Delta^0   \quad \text{and}\quad\
\psi^{slm}\cong \Delta^{-s}
\end{equation}
%~~~~~~~~~~~~~~~~~~~~~~~~~~~~~~~~~~~~~~~~~~~~~~~~~~~~~~~~~~~~~~~~~~~~~~
(to the leading order in $\Delta$).
However, it is also possible to write an exact, global, basis of
solutions to the static equation, as we do now.

In terms of the radial variable $z$,
Eq.\ (\ref{eq10}) takes the form
%~~~~~~~~~~~~~~~~~~~~~~~~~~~~~~~~~~~~~~~~~~~~~~~~~~~~~~~~~~~~~~~~~~~~~~
\begin{equation} \label{eq10z}
(-z)(1+z)F_{0}''+\left[-(s+1)-2(s+1)z\right]F_{0}'+(l-s)(l+s+1)F_0=0,
\end{equation}
%~~~~~~~~~~~~~~~~~~~~~~~~~~~~~~~~~~~~~~~~~~~~~~~~~~~~~~~~~~~~~~~~~~~~~~
where a prime denotes $d/dz$.
This is the {\em hypergeometric equation} \cite{hg} for $F_0(-z)$.
One solution for Eq.\ (\ref{eq10z}) is given by
(see Sec.\ 2.1.1 in \cite{hg})
%~~~~~~~~~~~~~~~~~~~~~~~~~~~~~~~~~~~~~~~~~~~~~~~~~~~~~~~~~~~~~~~~~~~~~~
\begin{equation} \label{eq11}
\phi_r(z)= \left\{
\begin{array}{ll}
F(-l+s,l+s+1;s+1;-z)\equiv \phi_{r}^{+} & \text{(for $s>0$)}, \\
(4M^2z)^{-s}F(-l,l+1;-s+1;-z)\equiv \phi_{r}^{-} & \text{(for $s<0$)},
\end{array}\right.
\end{equation}
%~~~~~~~~~~~~~~~~~~~~~~~~~~~~~~~~~~~~~~~~~~~~~~~~~~~~~~~~~~~~~~~~~~~~~~
where $F$ denotes the hypergeometric function. (Hereafter we
often omit the indices $slm$ for brevity). Note that since
in both cases the first index is a non-positive integer,
$F$ is simply a {\em polynomial} in $z$, and so is $\phi_r$.
(We choose this notation because, as we shall see below, $\phi_r$ is
physically regular at the EH, whereas the other static
solution, to which we shall later refer as $\phi_{ir}$, is
irregular there.)
The normalization in Eq.\ (\ref{eq11}) was chosen so as to conform
with Eq.\ (\ref{eq10a}) [recall that at $z=0$, the hypergeometric
function $F=1$, and note also the relation $\Delta=4M^2 z(z+1)$].
Thus, to the leading order in $\Delta$, $\phi_r$ is given by
%~~~~~~~~~~~~~~~~~~~~~~~~~~~~~~~~~~~~~~~~~~~~~~~~~~~~~~~~~~~~~~~~~~~~~~
\begin{equation} \label{eq12}
\phi_r(r)\cong \left\{
\begin{array}{ll}
\Delta^0 & \text{(for $s>0$)}, \\
\Delta^{-s} & \text{(for $s<0$)}.
\end{array}\right.
\end{equation}
%~~~~~~~~~~~~~~~~~~~~~~~~~~~~~~~~~~~~~~~~~~~~~~~~~~~~~~~~~~~~~~~~~~~~~~

A second, independent, static solution is given by
(see Sec.\ 2.2.2, case 21, in \cite{hg})
%~~~~~~~~~~~~~~~~~~~~~~~~~~~~~~~~~~~~~~~~~~~~~
\begin{equation} \label{eq13}
\phi_{ir}(z)= A_{ls}\times \left\{
\begin{array}{ll}
(4M^2z)^{-s}(1+z)^{-l-1}F[l-s+1,l+1;2l+2;(1+z)^{-1}]
\equiv \phi_{ir}^{+}                         & \text{(for $s>0$)}, \\
(1+z)^{-l-s-1}F[l+s+1,l+1;2l+2;(1+z)^{-1}]
\equiv \phi_{ir}^{-}                         & \text{(for $s<0$)},
\end{array}\right.
\end{equation}
%~~~~~~~~~~~~~~~~~~~~~~~~~~~~~~~~~~~~~~~~~~~~~
where $A_{sl}$ is a normalization factor,
%~~~~~~~~~~~~~~~~~~~~~~~~~~~~~~~~~~~~~~~~~~~~~
\begin{equation} \label{eq14}
A_{sl}=1/F(l-|s|+1,l+1;2l+2;1)=\frac{l!(l+|s|)!}{(2l+1)!(|s|-1)!}
\end{equation}
%~~~~~~~~~~~~~~~~~~~~~~~~~~~~~~~~~~~~~~~~~~~~~
(cf.\ Eq.\ (46) in Sec.\ 2.8 of \cite{hg}), chosen such that
$\phi_{ir}$ takes the simple leading-order asymptotic form
(\ref{eq10a}) at the EH, namely
%~~~~~~~~~~~~~~~~~~~~~~~~~~~~~~~~~~~~~~~~~~~~~~~~~~~~~~~~~~~~~~~~~~~~~~
\begin{equation} \label{eq12d}
\phi_{ir}(r)\cong \left\{
\begin{array}{ll}
\Delta^{-s} & \text{(for $s>0$)}, \\
\Delta^{0} & \text{(for $s<0$)}.
\end{array}\right.
\end{equation}
%~~~~~~~~~~~~~~~~~~~~~~~~~~~~~~~~~~~~~~~~~~~~~~~~~~~~~~~~~~~~~~~~~~~~~~

A careful study of the asymptotic behavior of $\phi_{ir}$ at the EH
reveals that it includes a (sub-dominant) logarithmic term.\footnote{
Such a logarithmic term is to be anticipated, because of the integer
difference, $|s|$, between the leading powers of $z$ in the two
asymptotic solutions (\ref{eq10a}) near the regular-singular point
$z=0$.\cite{Wayland}
}
To analyze this logarithmic term, it is instructive to express the
irregular solution in terms of $\phi_{r}$ via the Wronskian method. The
Wronskian associated with the homogeneous equation
(\ref{eq10}) is
%~~~~~~~~~~~~~~~~~~~~~~~~~~~~~~~~~~~~~~~~~~~~~
\begin{equation} \label{W}
W=\Delta^{-s-1},
\end{equation}
%~~~~~~~~~~~~~~~~~~~~~~~~~~~~~~~~~~~~~~~~~~~~~
and thus a static solution independent of $\phi_{r}$ may be expressed
as
%~~~~~~~~~~~~~~~~~~~~~~~~~~~~~~~~~~~~~~~~~~~~~
\begin{eqnarray} \label{integral}
\tilde{\phi}_{ir}&=&
-\phi_r(r) \int^r \phi_r^{-2}(r')W(r')dr'       \nonumber\\
&=&-\phi_r(r) \int^r \phi_r^{-2}(r')[\Delta(r')]^{-s-1}dr'.
\end{eqnarray}
%~~~~~~~~~~~~~~~~~~~~~~~~~~~~~~~~~~~~~~~~~~~~~
This solution, of course, does not necessarily coincide with $\phi_{ir}$,
but is, in general, a linear combination of the two basis functions
$\phi_r$ and $\phi_{ir}$.
It is easy to verify that the integrand in Eq.\ (\ref{integral})
is $z^{-|s|-1}$ times a rational function which is regular
(and nonvanishing) at $z=0$. The integrand can therefore be expanded as
%~~~~~~~~~~~~~~~~~~~~~~~~~~~~~~~~~~~~~~~~~~~~~
\begin{equation} \label{integrand}
z^{-|s|-1} (\gamma_0+\gamma_1 z+\gamma_2 z^2+\cdots),
\end{equation}
%~~~~~~~~~~~~~~~~~~~~~~~~~~~~~~~~~~~~~~~~~~~~~
where $\gamma_i$ are constants, with $\gamma_0 =(4M^2)^{-|s|-1}\neq 0$.
By substituting the leading-order term of this expansion in Eq.\
(\ref{integral}) and comparing to Eq.\ (\ref{eq12d}), we find
%~~~~~~~~~~~~~~~~~~~~~~~~~~~~~~~~~~~~~~~~~~~~~
\begin{equation} \label{combination}
\tilde{\phi}_{ir}= (2M|s|)^{-1}\phi_{ir}+ {\rm const}\cdot\phi_r.
\end{equation}
%~~~~~~~~~~~~~~~~~~~~~~~~~~~~~~~~~~~~~~~~~~~~~
[Here, the coefficient of $\phi_r$ depends on the specific choice
of the lower integration limit in Eq.\ (\ref{integral}).]
Substitution of the full expansion (\ref{integrand}) in
Eq.\ (\ref{integral}) yields
%~~~~~~~~~~~~~~~~~~~~~~~~~~~~~~~~~~~~~~~~~~~~~
\begin{equation} \label{integral1}
\tilde{\phi}_{ir} = -\phi_r\left[
\gamma_{|s|}\ln z+ z^{-|s|} (\hat \gamma_0+\hat\gamma_1 z+\cdots)
\right],
\end{equation}
%~~~~~~~~~~~~~~~~~~~~~~~~~~~~~~~~~~~~~~~~~~~~~
where $\hat \gamma_i =2M \gamma_i/(i-|s|)$ for $i\neq |s|$
(and $\hat \gamma_{|s|}$ is an arbitrary integration constant).
We now use Eq.\ (\ref{combination}) to extract $\phi_{ir}$:
%~~~~~~~~~~~~~~~~~~~~~~~~~~~~~~~~~~~~~~~~~~~~~
\begin{eqnarray} \label{combination1}
\phi_{ir}&=& 2M|s|(\tilde\phi_{ir}- {\rm const}\cdot\phi_r)
                                                    \nonumber\\
&=& -2M|s|\,\phi_r\left[\gamma_{|s|}\ln z+ z^{-|s|}
(\hat \gamma_0+\hat\gamma_1 z+\cdots) + {\rm const} \right].
\end{eqnarray}
%~~~~~~~~~~~~~~~~~~~~~~~~~~~~~~~~~~~~~~~~~~~~~
It is straightforward to expand this expression about $z=0$.
This expansion yields
%~~~~~~~~~~~~~~~~~~~~~~~~~~~~~~~~~~~~~~~~~~~~~
\begin{equation} \label{eq45}
\phi_{ir}(r)= \left\{
\begin{array}{ll}
\Delta^{-s}(1+\alpha^{+}_{1}z+\alpha^{+}_{2}z^2+\cdots) +
\beta_{sl}\,\phi_r^+\ln z
                                        & \text{(for $s>0$)}, \\
(1+\alpha^{-}_{1}z+\alpha^{-}_{2}z^2+\cdots) +
\beta_{sl}\,\phi_r^-\ln z
                                        & \text{(for $s<0$)},
\end{array}\right.
\end{equation}
%~~~~~~~~~~~~~~~~~~~~~~~~~~~~~~~~~~~~~~~~~~~~~
where $\alpha_{i}^{\pm}$ and $\beta_{sl}=-2M|s|\gamma_{|s|}$
are constants.

The above analysis, based on Eq.\ (\ref{integral}), explains the origin of
the logarithmic term and determines its exact form. The calculation of
$\gamma_{|s|}$ (and hence of $\beta_{sl}$) for general $l,s$ is tedious,
however. It is easier to derive the explicit expression for $\beta_{sl}$
directly from the exact expression (\ref{eq13}) for $\phi_{ir}$. The
series expansion of the hypergeometric function around $(1+z)^{-1}=1$
(corresponding to $z=0$) may be obtained from a generating function
through the formula
%~~~~~~~~~~~~~~~~~~~~~~~~~~~~~~~~~~~~~~~~~~~~~
\begin{equation} \label{eq15}
F(l-|s|+1,l+1,2l+2,y)=\frac{(-1)^{|s|+1}(2l+1)!}{(l!)^2(l-|s|)!(l+|s|)!}
\frac{d^l}{dy^l}\left[(1-y)^{l+|s|}\frac{d^l}{dy^l}
\left(\frac{\ln(1-y)}{y}\right)\right]
\end{equation}
%~~~~~~~~~~~~~~~~~~~~~~~~~~~~~~~~~~~~~~~~~~~~~
(cf.\ Eq.\ (4) in Sec.\ 2.2.2 of \cite{hg}).
Note that in our case $y=1/(1+z)$, so $\ln(1-y)=\ln z-\ln(1+z)$.
The logarithmic term in Eq.\ (\ref{eq45}) (which comes from the
first of the above two $\ln$ terms) is only obtained when none of the
$2l$ derivative operators $d/dy$ in Eq.\ (\ref{eq15}) acts on
$\ln(1-y)$. Thus, for the sake of calculating the logarithmic coefficient,
we can replace the second factor in the right-hand side of Eq.\
(\ref{eq15}) by
%~~~~~~~~~~~~~~~~~~~~~~~~~~~~~~~~~~~~~~~~~~~~~
\begin{equation} \label{eq15a}
\frac{d^l}{dy^l}\left[(1-y)^{l+|s|}\frac{d^l}{dy^l}
(y^{-1})\right] \ln z
=(-1)^l (l)! \frac{d^l}{dy^l}\left[(1-y)^{l+|s|}y^{-1-l}\right]
\ln z.
\end{equation}
%~~~~~~~~~~~~~~~~~~~~~~~~~~~~~~~~~~~~~~~~~~~~~
Evaluating this expression at $y=1$, to the leading order in
$z\cong 1-y$, we obtain
%~~~~~~~~~~~~~~~~~~~~~~~~~~~~~~~~~~~~~~~~~~~~~
\begin{equation} \label{eq15b}
\frac{(l)!(l+|s|)!}{(|s|)!}\, z^{|s|} \ln z.
\end{equation}
%~~~~~~~~~~~~~~~~~~~~~~~~~~~~~~~~~~~~~~~~~~~~~
Substituting this in Eq.\ (\ref{eq15}), and recalling Eq.\ (\ref{eq14}),
we obtain the desired expression for the logarithmic coefficient:
%~~~~~~~~~~~~~~~~~~~~~~~~~~~~~~~~~~~~~~~~~~~~~
\begin{equation} \label{eq46}
\beta_{sl}=\frac{(-1)^{s+1}(l+|s|)!}{(|s|-1)!(|s|)!(l-|s|)!}
\, (4M^2)^{-|s|}
\neq 0.
\end{equation}
%~~~~~~~~~~~~~~~~~~~~~~~~~~~~~~~~~~~~~~~~~~~~~

In summary, we have constructed a basis of solutions
to the static field equation. One of the basis solutions
($\phi_r$) is simply a polynomial in $z$, but the other
($\phi_{ir}$) contains a logarithmic term. This logarithmic term
will play an important role in the analysis below.
Note also that for both $s>0$ and $s<0$, $\phi_r$ is smaller
than $\phi_{ir}$ in the leading order by a factor $\Delta ^{|s|}$.

\section{Regularity at the EH} \label{regularity}

By general considerations, we expect physical
perturbations to be regular and smooth at the EH. The
function $\Psi^s$ represents a perturbation in the Maxwell
field tensor $F_{\alpha \beta }$ for $s=\pm 1$ and in the
Weyl tensor $C_{\alpha \beta \gamma \delta }$ for $s=\pm
2$. When this perturbation is expressed in Kruskal
coordinates (\ref{Kruskal}) (or in any other coordinates which are
regular at the EH), all components of these Maxwell or Weyl tensors
must take a perfectly regular form at the EH.

To discuss the regularity of $\Psi^s$ at the EH, it is
useful to define
%~~~~~~~~~~~~~~~~~~~~~~~~~~~~~~~~~~~~~~~~~~~~~~~~~~~~~~~~~~~~~~~~~~~~~~
\begin{equation} \label{eq51}
\hat \Psi^s\equiv \Delta^s\,\Psi^s,
\end{equation}
%~~~~~~~~~~~~~~~~~~~~~~~~~~~~~~~~~~~~~~~~~~~~~~~~~~~~~~~~~~~~~~~~~~~~~~
and correspondingly,
%~~~~~~~~~~~~~~~~~~~~~~~~~~~~~~~~~~~~~~~~~~~~~~~~~~~~~~~~~~~~~~~~~~~~~~
\begin{equation} \label{eq52}
\hat \psi^{slm}\equiv \Delta^s\,\psi^{slm} \;\;\;\;\;\;
\text{and}\;\;\;\;\;\;
\hat F_{k}^{slm}\equiv \Delta^s\,F_{k}^{slm}.
\end{equation}
%~~~~~~~~~~~~~~~~~~~~~~~~~~~~~~~~~~~~~~~~~~~~~~~~~~~~~~~~~~~~~~~~~~~~~~
It is straightforward to show, via Eqs.\ (\ref{Psi}) and (\ref{varphi}),
that for any $s$, $\hat\Psi^s$ directly represents a physical
perturbation field which
must be regular at the EH: For $s=2$, $\hat\Psi^s$ is a
linear combination of Weyl components $C_{VaVb}$, where the
indices $a,b$ represent the two angular coordinates $\theta
,\varphi $. For $s=-2$, $\hat\Psi^s$ is a linear
combination of Weyl components $C_{UaUb}$. Similarly, for
$s=1$ and $s=-1$ $\hat\Psi^s$ is a linear combination of
Maxwell components $F_{Va}$ and $F_{aU}$, respectively.
Therefore, a necessary condition for
regularity at the EH is that $\hat\Psi^s$ be regular
(i.e. finite and smooth). Since the spin-weighted spherical
harmonics are smooth, $\hat\psi^{slm}$ must be smooth too.

We point out that the regularity of $\hat\Psi^s$ is
also dictated by mathematical considerations, as follows:
If one transforms the master equation
(\ref{Master}) from $\Psi^s$ to $\hat\Psi^s$, and from the
original coordinates to the Kruskal coordinates (\ref{Kruskal}),
the field equation becomes perfectly regular at the EH
(whereas with the original dependent variable $\Psi^s$ the equation
is singular at the EH, even in Kruskal coordinates).
Therefore, from the hyperbolic nature of the field
equation, if the initial data for $\hat\Psi^s$ are
regular (which we assume), no irregularity may evolve at
the EH.

Consider next the regularity of the static solutions.
We assume that for any $s$ and any $l,m$, there exists (at
least) one static solution which is physically regular at
the EH. For, if there is an external static source of a
multipole $l,m$ (and no incoming waves from past null
infinity), the field outside the BH will be static; and we
do expect this static field to be regular at the EH. The
presence of two independent regular static solutions (for a
given $s,l,m$) at the EH would violate the no-hair principle,
because then {\it all} static solutions would be regular at
the EH, including the one which is regular at infinity. We
shall now show, however, that for any $s$, one of the static
solutions (the solution $\phi_{ir}$) is irregular.

For $s<0$, the irregularity of $\phi_{ir}$ is obvious,
because the corresponding field
$\hat\phi_{ir}\equiv\Delta^s \phi_{ir}$ diverges
like $\Delta ^s$. For $s>0$ the field $\hat\phi_{ir}$ is
finite ($\cong \Delta ^0$) at the EH. Yet, the
logarithmic term implies that the solution is not smooth:
The derivative of order $|s|$ with respect to $r$ (which itself
is a regular coordinate) diverges. On the other hand, for
both $s>0$ and $s<0$, the field $\hat\phi_r\equiv\Delta^s\phi_r$
is a polynomial in $z$ [which is proportional to
$(r-2M)^s$ for $s>0$ and to $(r-2M)^0$ for $s<0$], so it is
perfectly smooth.

We conclude that for both positive and negative $s$,
$\phi_{ir}$ is physically irregular, whereas $\phi_r$ is
physically regular.

\section{Late-time behavior} \label{behavior}

The demand for regularity of $\hat\psi^{slm}$ at the EH
has immediate implications to the late-time expansion (\ref{eq4}).
Since $r$ and $v$ form a regular coordinate system for the
Schwarzschild background (the so-called ``ingoing Eddington
coordinates''), $\hat\psi^{slm}$ must be a perfectly smooth
function of $r$ and $v$ at the EH (recall that $r$ and $v$ are
related to $U$ and $V$ by an invertible analytic transformation).
Therefore, for any $k$ (and any $s$), $\hat
F_k^{slm}(r)\equiv \Delta ^s F_k^{slm}(r)$
must be a smooth function of $r$:
%~~~~~~~~~~~~~~~~~~~~~~~~~~~~~~~~~~~~~~~~~~~~~~~~~~~~~~~~~~~~~~~~~~~~~~
\begin{equation} \label{eq53}
\hat F_k^{slm}(r)\in C^{\infty}(\mathbb{R}) \quad \text{for all $k$}.
\end{equation}
%~~~~~~~~~~~~~~~~~~~~~~~~~~~~~~~~~~~~~~~~~~~~~~~~~~~~~~~~~~~~~~~~~~~~~~

In section \ref{LTE} we have shown that $F_{k=0}^{slm}$
must be a static solution. The regularity of $\hat F_{k=0}^{slm}(r)$
then implies that $F_{k=0}^{slm}$ must
coincide (up to some multiplicative constant) with the {\em
regular} static solution $\phi_r$. Hence, to the leading
order in $1/v$, we obtain
%~~~~~~~~~~~~~~~~~~~~~~~~~~~~~~~~~~~~~~~~~~~~~~~~~~~~~~~~~~~~~~~~~~~~~~
\begin{equation} \label{eq55}
\psi^{slm}(t,r)=c_0\,\phi_r(r)\,v^{-2l-3}+O(v^{-2l-4}),
\end{equation}
%~~~~~~~~~~~~~~~~~~~~~~~~~~~~~~~~~~~~~~~~~~~~~~~~~~~~~~~~~~~~~~~~~~~~~~
where $c_0$ is constant.
For the description of the late-time behavior along
worldlines of fixed $r>2M$, it is useful to re-write this
expression in terms of powers of $1/t$:\footnote{
It should be stressed here that in this paper we {\em assume} the
power index $2l+3$ derived by Price \cite{PriceII} for the tail at fixed
$r>2M$. The new information in Eq.\ (\ref{eq56}) [or in Eq.\ (\ref{eq55})]
concerns the explicit form of the radial function multiplying the
inverse-power factor.}
%~~~~~~~~~~~~~~~~~~~~~~~~~~~~~~~~~~~~~~~~~~~~~~~~~~~~~~~~~~~~~~~~~~~~~~
\begin{equation} \label{eq56}
\psi ^{slm}(r,t)=c_0\,\phi_r(r)\,t^{-2l-3}+O(t^{-2l-4}).
\end{equation}
%~~~~~~~~~~~~~~~~~~~~~~~~~~~~~~~~~~~~~~~~~~~~~~~~~~~~~~~~~~~~~~~~~~~~~~

From this point on we discuss the cases $s<0$ and $s>0$
separately.

\subsection{The case $s<0$}

In this case, $F_{k=0}^{slm}$ is proportional to $\phi_r^-$.
We shall denote the proportionality constant by $c_0^-$,
that is,
%~~~~~~~~~~~~~~~~~~~~~~~~~~~~~~~~~~~~~~~~~~~~~~~~~~~~~~~~~~~~~~~~~~~~~~
\begin{equation} \label{eq57}
F_{k=0}^{slm}=c_0^-\,\phi _r^-\quad (s<0).
\end{equation}
%~~~~~~~~~~~~~~~~~~~~~~~~~~~~~~~~~~~~~~~~~~~~~~~~~~~~~~~~~~~~~~~~~~~~~~
(Recall that the parameter $k_0$ in Eq.\ (\ref{eq4}) is so defined
such that the term $F^{slm}_{k=0}$ does not vanish identically.
Therefore, by definition, the constant $c_0^-$ in non-zero.)
Note that $\phi_r^-\cong \Delta ^{|s|}$ near the EH.

Consider next the contribution from the terms $k>0$.
From Eq.\ (\ref{eq53}) it is obvious that
for $s<0$ and for all $k>0$, $F_{k}^{slm}$ must be a regular function
of $r$, which vanishes at least like $\Delta^{|s|}$ at the EH
(like for $k=0$).
Hence, at late time the terms $k>0$ are negligible compared to the
term $k=0$, due to their higher negative powers of $1/v$.
Therefore, Eq.\ (\ref{eq55}), which now reads
%~~~~~~~~~~~~~~~~~~~~~~~~~~~~~~~~~~~~~~~~~~~~~~~~~~~~~~~~~~~~~~~~~~~~~~
\begin{equation} \label{eq58}
\psi^{slm}(r,t)\cong c_0^-\,\phi _r^-(r)\,v^{-2l-3}
\quad \quad (s<0),
\end{equation}
%~~~~~~~~~~~~~~~~~~~~~~~~~~~~~~~~~~~~~~~~~~~~~~~~~~~~~~~~~~~~~~~~~~~~~~
provides a useful description of the late-time behavior not
only at $r>2M$ but also at the EH. To the leading order in
$\Delta$, the asymptotic behavior at the EH is
%~~~~~~~~~~~~~~~~~~~~~~~~~~~~~~~~~~~~~~~~~~~~~~~~~~~~~~~~~~~~~~~~~~~~~~
\begin{equation} \label{eq59}
\psi ^{slm}(r,t)\cong c_0^- \,\Delta ^{-s}\,v^{-2l-
3}\quad \quad (s<0).
\end{equation}
%~~~~~~~~~~~~~~~~~~~~~~~~~~~~~~~~~~~~~~~~~~~~~~~~~~~~~~~~~~~~~~~~~~~~~~

\subsection{The case $s>0$}

In this case, $F_{k=0}^{slm}$ (which must coincide with
the regular static solution) is proportional to $\phi_r^+$, i.e.,
$F_{k=0}^{slm}=c_0^+\phi_r^+$ with $c_0^+\neq 0$;
hence $F_{k=0}^{slm}\propto \Delta ^0$ near the EH.
However, for $k>0$ (for which $F_{k}^{slm}$ is not the static
solution), the only obvious constraint on the functions
$F_{k}^{slm}$ is the regularity of $\hat F_k^{slm}(r)$,
Eq.\ (\ref{eq53}).
This regularity criterion allows the terms $k>0$ to be proportional
to $\Delta ^{-s}$ (and, as we show below, at least the term $k=1$
is indeed proportional to $\Delta ^{-s}$).
Due to this $\Delta ^{-s}$ factor, at the EH the $O(v^{-2l-4})$ term in
Eq.\ (\ref{eq55}) dominates the $O(v^{-2l-3})$ term, which is only
proportional to $\Delta^0$. Therefore, for $s>0$,
Eq.\ (\ref{eq55}) does not provide a useful description of the
asymptotic behavior at the EH as it does for $r>2M$.
A correct description of the late-time
behavior there must include both terms $k=0$ and $k=1$:
%~~~~~~~~~~~~~~~~~~~~~~~~~~~~~~~~~~~~~~~~~~~~~~~~~~~~~~~~~~~~~~~~~~~~~~
\begin{equation} \label{eq61}
\psi^{slm}(r,t)=c_0^+\,\phi_r^+(r)\,v^{-2l-3} + F_{k=1}^{slm}(r)\,
v^{-2l-4} +O (v^{-2l-5})\quad \quad (s>0).
\end{equation}
%~~~~~~~~~~~~~~~~~~~~~~~~~~~~~~~~~~~~~~~~~~~~~~~~~~~~~~~~~~~~~~~~~~~~~~
To the leading order in $\Delta$, the asymptotic behavior
at the EH is
%~~~~~~~~~~~~~~~~~~~~~~~~~~~~~~~~~~~~~~~~~~~~~~~~~~~~~~~~~~~~~~~~~~~~~~
\begin{equation} \label{eq62}
\psi^{slm}(r,t)\cong c_0^+\,v^{-2l-3}+c_1^+\Delta ^{-s}\,
v^{-2l-4}\quad \quad (s>0),
\end{equation}
%~~~~~~~~~~~~~~~~~~~~~~~~~~~~~~~~~~~~~~~~~~~~~~~~~~~~~~~~~~~~~~~~~~~~~~
where $c_1^+$ is the coefficient of $\Delta^{-s}$ in
$F_{k=1}^{slm}$.
Note that Eqs.\ (\ref{eq55}) and (\ref{eq56}) still provide a correct
and useful description of the late-time behavior along any
line of fixed $r>2M$.

It is important to verify that the coefficient $c_1^+$
in Eq.\ (\ref{eq62}) is non-vanishing. This coefficient is to be
obtained from the function $F_{k=1}^{slm}(r)$ in Eq.\ (\ref{eq61}).
$F_{k=1}^{slm}$ satisfies the inhomogeneous equation (\ref{eq5}),
subject to the regularity condition (\ref{eq53}).
The general inhomogeneous solution takes the form
%~~~~~~~~~~~~~~~~~~~~~~~~~~~~~~~~~~~~~~~~~~~~~
\begin{equation} \label{eq63}
F_{k=1}^{slm}(r)=
a_1^+\phi_r^+(r) + b_1^+\phi_{ir}^+(r) +\phi_{ih}(r),
\end{equation}
%~~~~~~~~~~~~~~~~~~~~~~~~~~~~~~~~~~~~~~~~~~~~~
where $a_1^+$ and $b_1^+$ are constants and $\phi_{ih}$ is a specific
inhomogeneous solution.
Using the Wronskian function $W(r)$ given in Eq.\ (\ref{W}),
we can express $\phi_{ih}$ as
%~~~~~~~~~~~~~~~~~~~~~~~~~~~~~~~~~~~~~~~~~~~~~
\begin{equation} \label{eq63e}
\phi_{ih}(r)=
\phi_r(r) \int^r
\frac{\tilde{\phi}_{ir}(r')S_1(r')/\Delta(r')}{W(r')}\,dr'-
\tilde{\phi}_{ir}(r) \int^r
\frac{\phi_r(r')S_1(r')/\Delta(r')}{W(r')}\,dr'.
\end{equation}
%~~~~~~~~~~~~~~~~~~~~~~~~~~~~~~~~~~~~~~~~~~~~~
For $s>0$ it is convenient to re-express this inhomogeneous
solution in the form
%~~~~~~~~~~~~~~~~~~~~~~~~~~~~~~~~~~~~~~~~~~~~~
\begin{equation} \label{eq63a}
\phi_{ih}(r)=
\int^r \!\!\!dr' \int_{2M}^{r'} \!\!\!dr''\,
\frac{\phi_r^+(r)\phi_r^+(r'')}{[\phi_r^+(r')]^2}\,
\frac{W(r')}{W(r'')}\, \frac{S_1(r'')}{\Delta(r'')},
\end{equation}
%~~~~~~~~~~~~~~~~~~~~~~~~~~~~~~~~~~~~~~~~~~~~~
which is easily obtained from Eq.\ (\ref{eq63e}) by first
substituting for $\tilde{\phi}_{ir}$, using Eq.\ (\ref{integral}),
and then integrating the resulting expression by parts.
The form (\ref{eq63a}) is advantageous as it only involves the
homogeneous solution $\phi_r^+$, which has a simple polynomial
form.\footnote{
In Eq.\ (\ref{eq63a}) we have not specified the lower limit of
the integration over $r'$. Changing the value of this limit
amounts to adding a regular solution $\propto \phi_r^+$, which
is equivalent to re-defining the coefficient $a_1^+$ in Eq.\ (\ref{eq63}).
Note, however, that the choice $r'=2M$ as the lower integration
limit is forbidden, as the integral is not defined in this case.
}
The source term $S_1$ is to be calculated
from Eq.\ (\ref{eq7}) with $F_{k-1}(r)=F_0(r)=c_0^+\phi_r^+(r)$.
This yields, to the leading order in $\Delta$,
%~~~~~~~~~~~~~~~~~~~~~~~~~~~~~~~~~~~~~~~~~~~~~
\begin{equation} \label{eq65}
S_1(r)\cong 16 M^3 s k_0 c_0^+ \Delta^{-1}.
\end{equation}
%~~~~~~~~~~~~~~~~~~~~~~~~~~~~~~~~~~~~~~~~~~~~~
In view of Eqs.\ (\ref{eq12}), (\ref{W}), and (\ref{eq65}),
we find that the integrand in Eq.\ (\ref{eq63a}) is given, to the
leading order in $\Delta$, by
$16 M^3 s k_0 c_0^+\Delta^{s-1}(r'')/\Delta^{s+1}(r')$.
Performing the double integration, we obtain (to the leading order in
$\Delta$)\footnote{
Note that no $\ln^2 z,\ln^3 z\ldots$ terms arise from
the integration in Eq. (\ref{eq63a}):
For $s>0$ the integrand is actually a rational function of
$r''$, analytic at $r''=2M$. Hence, the integration over $r''$
cannot produce a $\ln z'$ term. A term $\propto \ln z$ arises
only from the subsequent integration over $r'$.
}
%~~~~~~~~~~~~~~~~~~~~~~~~~~~~~~~~~~~~~~~~~~~~~
\begin{equation} \label{eq63b}
\phi_{ih}(r)\cong 4M k_0 c_0^+\ln z +O(\Delta^0).
\end{equation}
%~~~~~~~~~~~~~~~~~~~~~~~~~~~~~~~~~~~~~~~~~~~~~
By substitution in the inhomogeneous equation (\ref{eq5}), one easily
verifies that the term $\ln z$ in $\phi_{ih}$ must be multiplied by a
homogeneous solution (otherwise the homogeneous operator $D^{sl}$,
acting on the logarithmic part of $\phi_{ih}$, would yield a term
proportional to $\ln z$ -- which cannot be balanced by the
logarithmic-free source term); and from Eq.\ (\ref{eq63b}) and Eqs.\
(\ref{eq12},\ref{eq12d}) it follows that this homogeneous solution must
be proportional to $\phi_r^+$. Therefore,
%~~~~~~~~~~~~~~~~~~~~~~~~~~~~~~~~~~~~~~~~~~~~~
\begin{equation} \label{eq63c}
\phi_{ih}(r)\cong
4M k_0 c_0^+\,\phi_r^+\ln z +O(\Delta^0)
\end{equation}
%~~~~~~~~~~~~~~~~~~~~~~~~~~~~~~~~~~~~~~~~~~~~~
(in which the $O(\Delta^0)$ term is logarithmic-free). We now
substitute this expression in Eq.\ (\ref{eq63}), using the asymptotic
forms (\ref{eq12}) and
(\ref{eq45}), and keeping only the leading order (proportional to
$\Delta^{-s}$) of the non-logarithmic part:
%~~~~~~~~~~~~~~~~~~~~~~~~~~~~~~~~~~~~~~~~~~~~~
\begin{equation} \label{eq66}
F_{k=1}^{slm}(r)\cong
b_1^+[\Delta^{-s}+\beta_{sl}\,\phi_r^+\ln z]
+ 4M k_0 c_0^+\,\phi_r^+\ln z.
\end{equation}
%~~~~~~~~~~~~~~~~~~~~~~~~~~~~~~~~~~~~~~~~~~~~~
Note that the coefficient $a_1^+$ in Eq.\ (\ref{eq63})
(which, in principle, is to be obtained by matching the solution
to the late-time field at null infinity \cite {Barack99II})
does not enter Eq.\ (\ref{eq66}), as $\phi_r^+$ includes neither
$\Delta^{-s}$ terms nor logarithmic terms.

Now, $F_1(r)$ must satisfy the regularity condition
(\ref{eq53}), so it cannot contain a logarithmic term.
This dictates the value of the constant $b_1^+$:
%~~~~~~~~~~~~~~~~~~~~~~~~~~~~~~~~~~~~~~~~~~~~~
\begin{equation} \label{eq67}
b_1^+= -4M k_0 c_0^+ \beta_{sl}^{-1} \neq 0.
\end{equation}
%~~~~~~~~~~~~~~~~~~~~~~~~~~~~~~~~~~~~~~~~~~~~~
One can now identify the non-vanishing coefficient $b_1^+$ with
the above leading order coefficient $c_1^+$ of $F_1(r)$ at the EH.
We conclude that the coefficient $c_1^+$ in Eq.\ (\ref{eq62}) is
non-vanishing.
As a consequence, we find that {\em on the EH itself} the perturbation
is dominated by the second term in Eq. (\ref{eq62}), and hence it
decays there like $v^{-2l-4}$.

In this section we have obtained the asymptotic form (\ref{eq62})
(and proved that the coefficient $c_1^+$ is non-vanishing) by
a direct analysis of $\psi^{slm}$ in the case $s>0$.
There is yet another way to obtain Eq.\ (\ref{eq62}), which,
being somewhat outside the main course of this paper, we describe
in detail in App.\ \ref{appA}:
It is well known that each single one of the perturbation fields
$s=1$ and $s=-1$ determines the
full electromagnetic perturbation, i.e.\ the  full Maxwell
tensor $F_{\alpha\beta}$ (up to a trivial addition of the
static Coulomb solution). Similarly, each of the
perturbation fields $s=2$ and $s=-2$ determines the full
gravitational perturbation, i.e.\ the perturbation in the
Weyl tensor (up to gauge, and up to a trivial addition of
the static multipoles with $l=0$ and $l=1$). In particular,
$\varphi_2$ determines $\varphi_0$, and $\Psi_4$
determines $\Psi_0$, and vice versa. We use this
fact in App. \ref{appA}, where we obtain the asymptotic behavior for
$s>0$ from that of $s<0$ (as we showed above, the latter is relatively
simple, because for $s<0$ the term $k=0$ in the late-time expansion
completely describes the late-time behavior at the EH). To that end,
we shall use in App.\ \ref{appA} the well known
{\em Teukolsky-Starobinsky identities}.

\section{Single Fourier modes} \label{monochromatic}

Consider a solution to the field equation (\ref{eq2}), having the form
%~~~~~~~~~~~~~~~~~~~~~~~~~~~~~~~~~~~~~~~~~~~~~~~~~~~~~~~~~~~~~~~~~~~~~~
\begin{equation} \label{eq81}
\psi(r,t) =\psi_\omega (r)\,e^{-i\omega t}.
\end{equation}
%~~~~~~~~~~~~~~~~~~~~~~~~~~~~~~~~~~~~~~~~~~~~~~~~~~~~~~~~~~~~~~~~~~~~~~
(The indices $s,l,m$, which, in fact, characterize both functions
$\psi$ and $\psi_{\omega}$, are omitted here and below for brevity.)
Each Fourier mode $\psi_\omega (r)$ then satisfies an ordinary
equation which may be written as
%~~~~~~~~~~~~~~~~~~~~~~~~~~~~~~~~~~~~~~~~~~~~~~~~~~~~~~~~~~~~~~~~~~~~~~
\begin{equation} \label{eq81a}
\frac{d^2}{dr_*^2}\left(r\Delta^{s/2}\psi_{\omega}\right)
+\left[\omega^2+i\omega s R(r)+V_{ls}(r)\right]
\left(r\Delta^{s/2}\psi_{\omega}\right)=0,
\end{equation}
%~~~~~~~~~~~~~~~~~~~~~~~~~~~~~~~~~~~~~~~~~~~~~~~~~~~~~~~~~~~~~~~~~~~~~~
in which $R(r)$ and $V_{ls}(r)$ are certain radial functions.
The asymptotic form of this equation near the horizon is
%~~~~~~~~~~~~~~~~~~~~~~~~~~~~~~~~~~~~~~~~~~~~~~~~~~~~~~~~~~~~~~~~~~~~~~
\begin{equation} \label{eq82}
\frac{d^2}{dr_*^2}\left(\Delta^{s/2}\psi_{\omega}\right)\cong
\left(s/4M+i\omega\right)^2
\left(\Delta^{s/2}\psi_{\omega}\right).
\end{equation}
%~~~~~~~~~~~~~~~~~~~~~~~~~~~~~~~~~~~~~~~~~~~~~~~~~~~~~~~~~~~~~~~~~~~~~~
The two asymptotic solutions at the EH are
%~~~~~~~~~~~~~~~~~~~~~~~~~~~~~~~~~~~~~~~~~~~~~~~~~~~~~~~~~~~~~~~~~~~~~~
\begin{equation} \label{eq83}
\psi_{\omega}^{a}\cong \Delta ^0e^{i\omega r_*}\quad\text{and}\quad
\psi_{\omega}^{b}\cong \Delta ^{-s}e^{-i\omega r_*}
\end{equation}
%~~~~~~~~~~~~~~~~~~~~~~~~~~~~~~~~~~~~~~~~~~~~~~~~~~~~~~~~~~~~~~~~~~~~~~
(where use has been made of the asymptotic relation
$e^{r_*/(4M)}\propto \Delta^{1/2}$).
At the limit $\omega \to 0$, these two asymptotic solutions
approach the two asymptotic static solutions, Eq.\ (\ref{eq10a})
-- just as one would expect.
We shall now show, however, that for $s>0$ the role of regular
and irregular solutions is interchanged as the limit
$\omega \to 0$ is approached.

In the case $\omega \neq 0$, too, we expect
one of the two solutions to be regular and the
other one to be singular (for the same reasons as in the
static case). We now substitute $\psi^a$ and $\psi^b$ in
Eq.\ (\ref{eq81}), and construct the corresponding physical fields
$\hat \psi\equiv \Delta^s\,\psi$ (which, as was
discussed in section \ref{regularity}, should be regular functions
at the EH). We denote the functions $\hat\psi$ obtained from
$\psi_{\omega}^a$ and $\psi_{\omega}^b$ by $\hat\psi^a$ and
$\hat\psi^b$, respectively, and find
%~~~~~~~~~~~~~~~~~~~~~~~~~~~~~~~~~~~~~~~~~~~~~~~~~~~~~~~~~~~~~~~~~~~~~~
\begin{equation} \label{eq84}
\hat\psi^{a}\equiv \Delta^s\psi_{\omega}^{a}e^{-i\omega t}
\cong \Delta ^se^{-i\omega u},\quad \quad
\hat\psi^{b}\equiv \Delta^s\psi_{\omega}^{b}e^{-i\omega t}
\cong e^{-i\omega v}.
\end{equation}
%~~~~~~~~~~~~~~~~~~~~~~~~~~~~~~~~~~~~~~~~~~~~~~~~~~~~~~~~~~~~~~~~~~~~~~
Recall that $v$ is regular at the EH, but $u$ is not (as
the EH is a surface of finite $v$ but infinite $u$). This implies
that $\hat\psi^{b}$ is regular, but $\hat\psi^{a}$
is irregular. (For $s<0$, $\hat\psi^{a}$ diverges at the EH.
For $s>0$, $\hat\psi^{a}$ is finite, but its $s$-order
derivative with respect to $U$ is indeterminate at the EH, and
higher-order derivatives diverge there.) We conclude that
for $\omega\neq 0$ (and for both $s>0$ and $s<0$), $\psi_{\omega}^{b}$
is regular and $\psi_{\omega}^{a}$ is singular. (This is a well known
result; see \cite{Teukolsky73}.)

Let us now compare this situation to the static case,
Eqs.\ (\ref{eq12}) and (\ref{eq12d}).
For $s<0$, the classification into regular and
irregular solutions is preserved at the limit $\omega \to 0$.
However, for $s>0$, the regular and irregular solutions
switch role in this static limit!

\section{Scalar-field toy-model} \label{toy}

To better understand the exchange of regular and
singular solutions at the limit $\omega \to 0$ (for $s>0$),
it is instructive to consider a simple scalar-field toy-model.
Let $\Phi $ be a minimally-coupled, massless, Klein-Gordon test
field on the Schwarzschild background. We make
here the assumption that, in an appropriate gauge, the
late-time behavior of the electromagnetic four-potential
$A_\alpha $ and of the linear metric perturbation
$h_{\alpha \beta }$ is qualitatively the same as that of a
scalar field (this assumption is somewhat vague, especially
because of the gauge ambiguity. Note, however, that at
least for the behavior of metric perturbations along
$r={\rm const}>2M$ lines, this assumption is
verified in \cite{PriceI}.) Correspondingly, we would expect that
the components $F_{aV}$ of the Maxwell tensor -- which are
made of terms like $A_{a,V}$ -- will qualitatively behave
at the EH like $\Phi_{,V}$. For the same reason, we would
expect $F_{aU}$ to behave at the EH like $\Phi _{,U}$.
Recalling the way $\Psi^{s=\pm 1}$ is constructed from
$F_{\alpha \beta}$ by projection on the tetrad (\ref{tetrad})
[see Eq.\ (\ref{varphi})], one intuitively expects that
$\Psi^{s=\pm 1}$ will qualitatively behave as follows:
%~~~~~~~~~~~~~~~~~~~~~~~~~~~~~~~~~~~~~~~~~~~~~~~~~~~~~~~~~~~~~~~~~~~~~~
\begin{equation} \label{eq91}
\Psi ^{s=1}\propto \Delta ^{-1}\Phi _{,v}\equiv \tilde
\Psi ^{s=1},\quad \quad \Psi ^{s=-1}\propto \Phi _{,u}
\equiv \tilde \Psi ^{s=-1}.
\end{equation}
%~~~~~~~~~~~~~~~~~~~~~~~~~~~~~~~~~~~~~~~~~~~~~~~~~~~~~~~~~~~~~~~~~~~~~~
Similarly, for the case $|s|=2$, one expects
[in view of Eq.\ (\ref{Psi})] that
%~~~~~~~~~~~~~~~~~~~~~~~~~~~~~~~~~~~~~~~~~~~~~~~~~~~~~~~~~~~~~~~~~~~~~~
\begin{equation} \label{eq91a}
\Psi ^{s=2}\propto \Delta^{-2}\Phi _{;vv}\equiv
\tilde \Psi ^{s=2},\quad\quad \Psi ^{s=-2}
\propto \Phi _{;uu}\equiv \tilde \Psi ^{s=-2}.
\end{equation}
%~~~~~~~~~~~~~~~~~~~~~~~~~~~~~~~~~~~~~~~~~~~~~~~~~~~~~~~~~~~~~~~~~~~~~~
(For brevity, we shall focus the following discussion on the case
$|s|=1$. Similar arguments apply to $|s|=2$ as well.)\footnote
{A more sophisticated toy model
would be obtained by replacing $F_{\alpha \beta}$ or
$C_{\alpha \beta \gamma \delta }$ in the definitions of the
Newman-Penrose fields by $\Phi _{;\alpha\beta}$ or
$\Phi_{;\alpha \beta \gamma \delta }$, respectively. Here we
adopt a simpler toy-model which is easier to calculate.}

For any mode $l,m$ of $\Phi$, the two static solutions
take the asymptotic forms
%~~~~~~~~~~~~~~~~~~~~~~~~~~~~~~~~~~~~~~~~~~~~~~~~~~~~~~~~~~~~~~~~~~~~~~
\begin{equation} \label{eq92}
\Phi_r\cong 1 + O(\Delta )\quad \text{and}\quad
\Phi_{ir}\cong r_*
\end{equation}
%~~~~~~~~~~~~~~~~~~~~~~~~~~~~~~~~~~~~~~~~~~~~~~~~~~~~~~~~~~~~~~~~~~~~~~
near the EH (cf.\ \cite{PriceI}). Clearly, $\Phi_r$ is the regular
mode, while $\Phi_{ir}$ is singular (as $r_*\rightarrow -\infty$ at
the EH).
Let us denote the functions $\tilde \Psi^{s}$ which
correspond to the regular and singular modes by $\tilde
\Psi_r^{s}$ and $\tilde\Psi_{ir}^{s}$,
respectively. Recalling that in the static case $\partial_
v=-\partial_u=(1/2)\,d/dr*=[\Delta/(2r^2)]\,d/dr$, we find
for $|s|=1$:\footnote{
In deriving the asymptotic form for $\tilde\Psi_r^s$ it has been
assumed that $d\Phi_r/dr$ does not vanish at the EH.
This assumption is justified, as it is known \cite{Barack99II,Legendre}
that the EH-regular static scalar field $\Phi_r$ is nothing but
$P_l[(r-M)/M]$, the Legendre polynomial of order $l$ (up to a
multiplicative constant).
At the EH we then have $d\Phi_r/dr\propto dP_l/dr=l(l+1)/2$,
which does not vanish (except for $l=0$).
}
%~~~~~~~~~~~~~~~~~~~~~~~~~~~~~~~~~~~~~~~~~~~~~~~~~~~~~~~~~~~~~~~~~~~~~~
\begin{equation} \label{eq93}
\tilde\Psi_r^s\propto\left\{
    \begin{array}{ll}
    \Delta^0,    & s=+1\\
    \Delta^1,    & s=-1
    \end{array} \right.
,\;\;\;\;\;
\tilde\Psi_{ir}^s\propto\left\{
    \begin{array}{ll}
    \Delta^{-1}, & s=+1\\
    \Delta^0,    & s=-1
    \end{array} \right..
\end{equation}
%~~~~~~~~~~~~~~~~~~~~~~~~~~~~~~~~~~~~~~~~~~~~~~~~~~~~~~~~~~~~~~~~~~~~~~

Consider next a single Fourier mode (of a given $l,m,\omega$)
$\Phi =\Phi_\omega (r)\,e^{-i\omega t}$.
The two asymptotic solutions of the radial function at the EH are
obtained by substituting $s=0$ in Eq.\ (\ref{eq83}):
%~~~~~~~~~~~~~~~~~~~~~~~~~~~~~~~~~~~~~~~~~~~~~
\begin{equation} \label{eq94}
    \Phi_{\omega}^{a}\cong e^{i\omega r*}[1+O(\Delta)],\;\;\;\;\;
    \Phi_{\omega}^{b}\cong e^{-i\omega r*}[1+O(\Delta)].
\end{equation}
%~~~~~~~~~~~~~~~~~~~~~~~~~~~~~~~~~~~~~~~~~~~~~
These two radial functions correspond to the field configurations
%~~~~~~~~~~~~~~~~~~~~~~~~~~~~~~~~~~~~~~~~~~~~~
\begin{equation} \label{eq95}
    \Phi^a\equiv e^{-i\omega t}\,\Phi_{\omega}^{a}\cong e^{-i\omega
u}[1+O(\Delta )]
    ,\;\;\;\;\;
    \Phi^b\equiv e^{-i\omega t}\,\Phi_{\omega}^{b}\cong e^{-i\omega
v}[1+O(\Delta)].
\end{equation}
%~~~~~~~~~~~~~~~~~~~~~~~~~~~~~~~~~~~~~~~~~~~~~
Since $u$ diverges at the EH (but $v$ is regular), it is
obvious that $\Phi^b$ is the regular solution, while
$\Phi^a$ is singular.
We shall denote the functions $\tilde\Psi^{s}$ which correspond
to the regular and singular modes in Eq.\ (\ref{eq95}) by
$\tilde \Psi^a$ and $\tilde \Psi^b$, respectively.
Using Eq.\ (\ref{eq91}), we find for $|s|=1$:
%~~~~~~~~~~~~~~~~~~~~~~~~~~~~~~~~~~~~~~~~~~~~~
\begin{equation} \label{eq96}
\tilde\Psi^a\propto\left\{
    \begin{array}{ll}
    e^{-i\omega u}[1+O(\Delta )],   & s=+1\\
    e^{-i\omega u}[1+O(\Delta )],   & s=-1
    \end{array} \right.
,\;\;\;\;\;
\tilde\Psi^b\propto\left\{
    \begin{array}{ll}
    \Delta^{-1}\,e^{-i\omega v}[1+O(\Delta)], & s=+1\\
    \Delta\,e^{-i\omega v}[1+O(\Delta )],     & s=-1
    \end{array} \right..
\end{equation}
%~~~~~~~~~~~~~~~~~~~~~~~~~~~~~~~~~~~~~~~~~~~~~
(It is assumed here that the terms $O(\Delta)$ in Eq.\
(\ref{eq95}) are non-vanishing, and, moreover, that their
derivatives with respect to $r$ do not vanish at the EH.)
The construction of $\tilde \Psi^a$ and $\tilde \Psi^b$ ensures that the
$t$-dependence of both functions is simply $e^{-i\omega t}$.
Let us denote the radial parts of these two functions by
$\tilde \Psi_{\omega}^{a}(r)$ and $\tilde \Psi_{\omega}^{b}(r)$,
respectively; that is,
%~~~~~~~~~~~~~~~~~~~~~~~~~~~~~~~~~~~~~~~~~~~~~
\begin{equation} \label{eq96a}
\tilde \Psi^a(r,t) \equiv
e^{-i\omega t}\tilde\Psi_{\omega}^{a}(r)
,\;\;\;\;\;
\tilde \Psi^b(r,t) \equiv
e^{-i\omega t}\tilde\Psi_{\omega}^{b}(r).
\end{equation}
%~~~~~~~~~~~~~~~~~~~~~~~~~~~~~~~~~~~~~~~~~~~~~
For both cases $s=1$ and $s=-1$, one thus finds
%~~~~~~~~~~~~~~~~~~~~~~~~~~~~~~~~~~~~~~~~~~~~~
\begin{equation} \label{eq97}
\tilde\Psi_{\omega}^{a}\propto e^{i\omega r_*}[1+O(\Delta)],\quad\quad
\tilde\Psi_{\omega}^{b}\propto \Delta ^{-s}\,e^{-i\omega r_*}
[1+O(\Delta )].
\end{equation}
%~~~~~~~~~~~~~~~~~~~~~~~~~~~~~~~~~~~~~~~~~~~~~

A comparison of Eq.\ (\ref{eq93}) to Eqs.\ (\ref{eq12},\ref{eq12d}),
and of Eq.\ (\ref{eq97}) to Eq.\ (\ref{eq83}), reveals that
for both cases $\omega = 0$ and $\omega \neq 0$, and for both $s=1$
and $s=-1$, the actual asymptotic form of both the regular and singular
solutions agree with that obtained from the scalar-field toy-model.
In particular, in the case $s=-1$, at the limit $\omega \to 0$
the regular $\omega \neq 0$ solution $\tilde\Psi_{\omega}^{b}$
approaches the regular static solution $\tilde\Psi_r^{s=-1}$
(and the singular $\omega \neq 0$ solution $\tilde \Psi_{\omega}^{a}$
approaches the singular static static solution
$\tilde \Psi _{ir}^{s=-1}$), whereas in the case $s=+1$ the regular
and singular solutions interchange at the limit $\omega \to 0$.

Our toy model provides a simple intuitive explanation
for the difference in the role of the regular and singular
solutions in the static and $\omega \neq 0$ cases.
The key point is the relation between the two basis
solutions of the scalar field itself, i.e. Eqs.\ (\ref{eq92})
and (\ref{eq94},\ref{eq95}).
In the static case, there is a ``small solution''
$\Phi _r$ and a ``large solution'' $\Phi _{ir}$. Naturally,
the ``small solution'' is the regular one, and the ``large
solution'' is singular. On the other hand, in the case
$\omega \neq 0$ both radial solutions in Eq.\ (\ref{eq94}) are of
the same magnitude. In this case, the fundamental
difference between the two basis solutions is that, at the
leading order, one of them ($\Phi^{a}$) depends solely on $u$,
and the other one ($\Phi^{b}$) depends solely on $v$.
We can therefore refer to the two radial functions $\Phi_{\omega}^{a}$
and $\Phi_{\omega}^{b}$ as the ``$u$-solution'' and the ``$v$-solution'',
respectively. Since $v$ is regular at the EH and $u$ diverges, the
``$v$-solution'' $\Phi_{\omega}^{b}$ is regular and the ``$u$-solution''
is singular.

Now, the functions $\tilde\Psi^s$ (which presumably
represent the functions $\Psi^s$) are obtained in our toy
model by differentiating $\Phi$ with respect to $u$ or $v$
(depending on the sign of $s$). Consider first the $\omega \neq
0$ case (in which the two basis solutions are classified as
a ``$v$-solution'' and a ``$u$-solution''). When the operator
$\partial _v$ acts on $\Phi $, it naturally yields a large
outcome for the ``$v$-solution'', and a small outcome for the
``$u$-solution''. On the other hand, when the operator
$\partial _u$ is applied, it yields a small outcome for the
``$v$-solution'', and a large outcome for the ``$u$-solution''.
Since the ``$v$-solution'' is regular and the ``$u$-solution'' is
singular, we arrive at the following conclusion: For
$\tilde \Psi ^{s=-1}$ (which is associated with $\Phi_{,u}$),
the regular solution is the smaller of the two
basic solutions. However, for $\tilde \Psi ^{s=1}$ (which
is associated with $\Phi _{,v}$), the regular solution is
the {\em larger} of the two basic solutions.

On the other hand, in the static case we have a ``large
solution'' and a ``small solution'' (instead of a ``$v$-solution''
and a ``$u$ solution''). The differentiation of the ``large
solution'' with respect to either $u$ or $v$ yields a function
$\tilde \Psi ^s$ which is larger than that obtained from
the differentiation of the ``small solution''. Therefore, in
the static case, for both $s>0$ and $s<0$ the regular
solution is the smaller of the two basis solutions.

The interchange of the regular and singular $s>0$
solutions in the transition from $\omega \neq 0$ to $\omega=0$
may still look somewhat mysterious, because the limit
$\omega \to 0$ is a perfectly regular limit of the
differential equation (\ref{eq81a}). The mystery may again be
resolved with the aid of our scalar-field toy model.
Let us re-write the regular $\omega \neq 0$ solution for
$\Phi $ [Eq.\ (\ref{eq95})] in a somewhat more
explicit form,
%~~~~~~~~~~~~~~~~~~~~~~~~~~~~~~~~~~~~~~~~~~~~~~~~~~~~~~~~~~~~~~~~~~~~~~
\begin{equation} \label{eq101}
\Phi^b\cong e^{-i\omega v}[1+c(\omega)\Delta +O(\Delta ^2)].
\end{equation}
%~~~~~~~~~~~~~~~~~~~~~~~~~~~~~~~~~~~~~~~~~~~~~~~~~~~~~~~~~~~~~~~~~~~~~~
We assume that $c(\omega)$ is continuous and non-vanishing at the limit
$\omega \to 0$.
We now calculate $\tilde \Psi_{\omega}^{s=1}$ from this regular solution,
via Eq.\ (\ref{eq91}), keeping the leading-order in $\Delta$
separately for terms proportional to
$\omega$ and for terms proportional to $\omega^0$:
%~~~~~~~~~~~~~~~~~~~~~~~~~~~~~~~~~~~~~~~~~~~~~~~~~~~~~~~~~~~~~~~~~~~~~~
\begin{equation} \label{eq102}
\tilde\Psi^{s=1}\cong e^{-i\omega v}[-i\omega \Delta^{-1}+
(4M)^{-1}c(\omega )]\,\,[1+O(\Delta )].
\end{equation}
%~~~~~~~~~~~~~~~~~~~~~~~~~~~~~~~~~~~~~~~~~~~~~~~~~~~~~~~~~~~~~~~~~~~~~~
Restricting attention to the limit $\omega \to 0$ and to
the leading order in $\Delta$, we obtain
%~~~~~~~~~~~~~~~~~~~~~~~~~~~~~~~~~~~~~~~~~~~~~~~~~~~~~~~~~~~~~~~~~~~~~~
\begin{equation} \label{eq103}
\tilde\Psi^{s=1}\cong e^{-i\omega v}(-i\omega \Delta^{-1}+c_0),
\end{equation}
%~~~~~~~~~~~~~~~~~~~~~~~~~~~~~~~~~~~~~~~~~~~~~~~~~~~~~~~~~~~~~~~~~~~~~~
where
$c_0=(4M)^{-1}\mathop {\lim }\limits_{\omega \to 0}c(\omega )$.
Equation (\ref{eq103}) explains the change in the asymptotic form
of the regular $s=1$ solution from $\Delta ^{-1}$ in the case
$\omega \neq 0$ to $\Delta ^0$ in the case $\omega = 0$.

On the other hand, when the same calculation is
carried out for $s=-1$, one obtains from Eq.\ (\ref{eq91})
%~~~~~~~~~~~~~~~~~~~~~~~~~~~~~~~~~~~~~~~~~~~~~~~~~~~~~~~~~~~~~~~~~~~~~~
\begin{equation} \label{eq104}
\tilde\Psi^{s=-1}\cong -c_0\,\Delta \,e^{-i\omega v}
\end{equation}
%~~~~~~~~~~~~~~~~~~~~~~~~~~~~~~~~~~~~~~~~~~~~~~~~~~~~~~~~~~~~~~~~~~~~~~
for the regular solution (for small $\omega$). Thus, the
regular solution for $s=-1$ is proportional to $\Delta$ for
both $\omega \neq 0$ and $\omega = 0$.

Our toy model also allows us to obtain the late-time
behavior for both $s>0$ and $s<0$ directly from that of the
scalar field. The late-time behavior for (a mode $l,m$ of)
$\Phi $ near the EH is known to be \cite{Gundlach94I,Barack99II}
%~~~~~~~~~~~~~~~~~~~~~~~~~~~~~~~~~~~~~~~~~~~~~~~~~~~~~~~~~~~~~~~~~~~~~~
\begin{equation} \label{eq105}
\Phi^{l}\cong \Phi_r^{l}(r)\,v^{-2l-3},
\end{equation}
%~~~~~~~~~~~~~~~~~~~~~~~~~~~~~~~~~~~~~~~~~~~~~~~~~~~~~~~~~~~~~~~~~~~~~~
where the radial function $\Phi_r^{l}(r)$ is the regular
static solution for the mode $l,m$. Equation (\ref{eq91})
now yields at the EH
%~~~~~~~~~~~~~~~~~~~~~~~~~~~~~~~~~~~~~~~~~~~~~~~~~~~~~~~~~~~~~~~~~~~~~~
\begin{equation} \label{eq106}
\tilde \Psi ^{s=-1}\cong -\Delta \,\Phi _1\,v^{-2l-3}
\end{equation}
%~~~~~~~~~~~~~~~~~~~~~~~~~~~~~~~~~~~~~~~~~~~~~~~~~~~~~~~~~~~~~~~~~~~~~~
and
%~~~~~~~~~~~~~~~~~~~~~~~~~~~~~~~~~~~~~~~~~~~~~~~~~~~~~~~~~~~~~~~~~~~~~~
\begin{equation} \label{eq107}
\tilde \Psi ^{s=1}\cong \Phi _1\,v^{-2l-3}
                    -(2l+3)\Phi _0\,\Delta ^{-1}\,v^{-2l-4},
\end{equation}
%~~~~~~~~~~~~~~~~~~~~~~~~~~~~~~~~~~~~~~~~~~~~~~~~~~~~~~~~~~~~~~~~~~~~~~
where $\Phi _0=\Phi _r^{l}(r=2M)$, and
$\Phi_1=(8M^2)^{-1} (d\Phi _r^{l}/dr)_{r=2M}$.
Compare these results to Eqs.\ (\ref{eq59}) and (\ref{eq62}).

So far we have implemented the toy model for the case
$|s|=1$ only. The calculations in the case $|s|=2$ are
straightforward too, though they are somewhat more tedious.
We shall merely point out here that all the expressions we
have derived for $\tilde \Psi ^{s=\pm 1}$ are extendible to
$\tilde \Psi ^{s=\pm 2}$, and may be used to explain the
various features of $\Psi ^{s=\pm 2}$ --- e.g.\ the
asymptotic behavior of the regular and singular solutions
for both $\omega \neq 0$ and $\omega = 0$, and the
late-time behavior at the EH.
It should be emphasized that the late-time power index of $\tilde \Psi
^{s=2}$ at the EH is $2l+4$ (and not $2l+5$, which might naively be
anticipated due to the two $v$-derivatives in the definition of
this function).
The reason is that, the second-order covariant differentiation in Eq.\
(\ref{eq91a}) involves the differentiation of the affine connection. The
easiest way to evaluate $\tilde \Psi ^{s=2}$ is via the Kruskal
coordinates (which at the EH minimize the connection's effect). One
then finds that $\Phi _{;V}\propto v^{-2l-4}/V$, and the next
differentiation with respect to $V$ then yields, at the leading order,
$\Phi _{;VV}\propto v^{-2l-4}/V^2$, i.e. $\Phi _{;vv}\propto v^{-2l-4}$.

Finally, we point out that Eq.\ (\ref{eq103}), which
was derived within the framework of the scalar-field toy
model, may also be derived for the realistic field $\Psi^{s=1}$,
if Eq.\ (\ref{eq104}) is assumed, using the Teukolsky-Starobinsky
identities (the application of which is described in the Appendix).
More explicitly, let us write the asymptotic behavior of the monochromatic
$s=-1$ field at the EH, to the leading order in $\Delta$, as
%~~~~~~~~~~~~~~~~~~~~~~~~~~~~~~~~~~~~~~~~~~~~~~~~~~~~~~~~~~~~~~~~~~~~~~
\begin{equation} \label{eq111}
\Psi^{s=-1}\cong \,a(\omega )\Delta \,e^{-i\omega v},
\end{equation}
%~~~~~~~~~~~~~~~~~~~~~~~~~~~~~~~~~~~~~~~~~~~~~~~~~~~~~~~~~~~~~~~~~~~~~~
and assume that $a(\omega )$ is non-vanishing at the limit $\omega \to
0$. Then, applying the Teukolsky-Starobinsky identities,
one can easily obtain for the corresponding $s=+1$ field
%~~~~~~~~~~~~~~~~~~~~~~~~~~~~~~~~~~~~~~~~~~~~~~~~~~~~~~~~~~~~~~~~~~~~~~
\begin{equation} \label{eq112}
\Psi^{s=1}\propto e^{-i\omega v}(-i\omega \Delta ^{-1}+const\cdot
\Delta ^0)
\end{equation}
%~~~~~~~~~~~~~~~~~~~~~~~~~~~~~~~~~~~~~~~~~~~~~~~~~~~~~~~~~~~~~~~~~~~~~~
(for small $\omega$).

\section{A Kerr Black Hole}\label{Kerr}

The above analysis of the Schwarzschild case has immediate
implications to rotating black holes as well. In a forthcoming paper
\cite{Kerr} the late time expansion will systematically be applied to
the Kerr case, in order to determine the late-time behavior of external
perturbations.
Here, we shall use the above methods and considerations to derive the
decay rate of $s\neq 0$ fields along the Kerr EH (many of the details
are left to Ref.\ \cite{Kerr}).

In the Kerr case, the Master equation is fully separable
only in the frequency domain, by writing
%~~~~~~~~~~~~~~~~~~~~~~~~~~~~~~~~~~~~~~~~~~~~~
\begin{equation} \label{eq112a}
\Psi^{\omega slm}(t,r,\theta,\varphi)=S_{\omega}^{slm}(\theta)
\,e^{im\varphi} e^{-i\omega t}\psi^{\omega slm}(r),
\end{equation}
%~~~~~~~~~~~~~~~~~~~~~~~~~~~~~~~~~~~~~~~~~~~~~
where $(t,r,\theta,\varphi)$ are the Boyer-Lindquist coordinates,
and $S_{\omega}^{slm}(\theta)\,e^{im\varphi}$ are the spin-weighted
spheroidal harmonics \cite{Teukolsky73}.
The behavior of the radial function $\psi^{\omega slm}(r)$ is then
governed by the well known Teukolsky equation \cite{Teukolsky73}.
The two asymptotic solutions of this equation at the EH are
%~~~~~~~~~~~~~~~~~~~~~~~~~~~~~~~~~~~~~~~~~~~~~
\begin{equation} \label{eq113}
\psi_a^{\omega slm}(r)\cong \Delta^{0}  e^{i(\omega-m\Omega_+)r_*},
\;\;\;\;\;
\psi_b^{\omega slm}(r)\cong \Delta^{-s} e^{-i(\omega-m\Omega_+)r_*},
\end{equation}
%~~~~~~~~~~~~~~~~~~~~~~~~~~~~~~~~~~~~~~~~~~~~~
where
%~~~~~~~~~~~~~~~~~~~~~~~~~~~~~~~~~~~~~~~~~~~~~
\begin{equation} \label{eq114b}
\Delta \equiv r^2-2Mr+a^2,
\end{equation}
%~~~~~~~~~~~~~~~~~~~~~~~~~~~~~~~~~~~~~~~~~~~~~
$M$ and $a$ are, respectively, the mass and specific angular momentum
of the black hole, $r_*$ is defined by $dr_*/dr=(r^2+a^2)/\Delta$, and
%~~~~~~~~~~~~~~~~~~~~~~~~~~~~~~~~~~~~~~~~~~~~~
\begin{equation} \label{eq114}
\Omega_+ \equiv \frac{a}{2Mr_+},
\end{equation}
%~~~~~~~~~~~~~~~~~~~~~~~~~~~~~~~~~~~~~~~~~~~~~
with $r_+\equiv M+(M^2-a^2)^{1/2}$ being the $r$-value of the EH.
[Compare the asymptotic solutions (\ref{eq113}) and (\ref{eq83})
in the case $a=0$.]
We shall consider only a BH background with $0<|a|<M$ (the extremal
case, $|a|=M$, requires a separate treatment).

In the Kerr case, the coordinate $\varphi$ goes singular at the EH.
Transforming to the regularized azimuthal coordinate
%~~~~~~~~~~~~~~~~~~~~~~~~~~~~~~~~~~~~~~~~~~~~~
\begin{equation} \label{eq114d}
\tilde\varphi_+\equiv \varphi-\Omega_+t
\end{equation}
%~~~~~~~~~~~~~~~~~~~~~~~~~~~~~~~~~~~~~~~~~~~~~
(see Sec.\ 58 in \cite{Chandra83}), and substituting the solutions
(\ref{eq113}) in Eq.\ (\ref{eq112a}), we obtain the field
configurations associated with the two asymptotic solutions:
%~~~~~~~~~~~~~~~~~~~~~~~~~~~~~~~~~~~~~~~~~~~~~
\begin{mathletters} \label{config}
\begin{equation} \label{eq114a1}
\Psi_a^{\omega slm}(t,r,\theta,\tilde\varphi_+)\cong
S_{\omega}^{slm}(\theta)\,e^{im\tilde\varphi_+}
\Delta^{0} e^{-i(\omega-m\Omega_+)u},
\end{equation}
\begin{equation} \label{eq114a2}
\Psi_b^{\omega slm}(t,r,\theta,\tilde\varphi_+)\cong
S_{\omega}^{slm}(\theta)\,e^{im\tilde\varphi_+}
\Delta^{-s} e^{-i(\omega-m\Omega_+)v},
\end{equation}
\end{mathletters}
%~~~~~~~~~~~~~~~~~~~~~~~~~~~~~~~~~~~~~~~~~~~~~
where
%~~~~~~~~~~~~~~~~~~~~~~~~~~~~~~~~~~~~~~~~~~~~~
\begin{equation} \label{eq112b}
\Psi_{a,b}^{\omega slm} \equiv S_{\omega}^{slm}(\theta)
\,e^{im\varphi} e^{-i\omega t}\psi_{a,b}^{\omega slm}(r).
\end{equation}
%~~~~~~~~~~~~~~~~~~~~~~~~~~~~~~~~~~~~~~~~~~~~~

It is straightforward to extend the regularity criterion of Sec.\  \ref
{regularity} to the Kerr case: Here, too, one finds that at the EH the
variable
%~~~~~~~~~~~~~~~~~~~~~~~~~~~~~~~~~~~~~~~~~~~~~
\begin{equation} \label{eq115}
\hat\Psi^s \equiv \Delta^s \Psi^s
\end{equation}
%~~~~~~~~~~~~~~~~~~~~~~~~~~~~~~~~~~~~~~~~~~~~~
must be a perfectly smooth function of the (regularized) coordinates
(exactly for the same reasons described in Sec.\  \ref {regularity} for
the Schwarzschild case; see also \cite{Teukolsky73}).

For the application of the late-time expansion we must verify which of
the above two asymptotic solutions is physically regular at the EH.
Teukolsky \cite{Teukolsky73} asserted that
the regular solution is $\Psi_b$. This is obvious from the
oscillatory
dependence of $\Psi_a$ on $u$ (and of $\Psi_b$ on $v$) --- as we have
discussed in the Schwarzschild ($w\neq 0$) case.
One must recall, however, that this
simple classification breaks down whenever $w-m\Omega_+=0$ (in which
case the above oscillatory factors in $u$ and $v$ degenerate to $1$).
In this case the classification is more involved.

We shall now restrict attention to the static\footnote{
Throughout this section, which deals with a Kerr
background, we refer to the $t$-independent solutions as "static", in a
slight abuse of the usual terminology. (We prefer to use here the term
"static" instead of "stationary" in order to simplify the terminology and
preserve the semantic analogy with the Schwarzschild case).
}
case $w=0$ (the case $w\neq 0$ is not required for the analysis below).
In this case, the asymptotic solutions (\ref{eq114a1},\ref{eq114a2})
become
%~~~~~~~~~~~~~~~~~~~~~~~~~~~~~~~~~~~~~~~~~~~~~
\begin{mathletters} \label{configstat}
\begin{equation} \label{eq114a10}
\Psi_a^{\omega=0, slm}(r,\theta,\tilde\varphi_+)\cong
Y^{slm}(\theta,\tilde\varphi_+)
\Delta^{0} e^{im\Omega_+ u},
\end{equation}
\begin{equation} \label{eq114a20}
\Psi_b^{\omega=0, slm}(r,\theta,\tilde\varphi_+)\cong
Y^{slm}(\theta,\tilde\varphi_+)
\Delta^{-s} e^{im\Omega_+ v},
\end{equation}
\end{mathletters}
%~~~~~~~~~~~~~~~~~~~~~~~~~~~~~~~~~~~~~~~~~~~~~
where $Y^{slm}$ denotes the spin-weighted spherical harmonics.
Teukolsky's assertion concerning the regularity of $\psi_b$ is now
valid for $m\neq 0$ only, and we still need to find out what is the
regular asymptotic behavior at the EH for $m=0$.

Fortunately, at this point we can directly apply the results from the
above analysis of the Schwarzschild case, for the following reason.
Let us define
%~~~~~~~~~~~~~~~~~~~~~~~~~~~~~~~~~~~~~~~~~~~~~
\begin{equation} \label{zKerr}
z\equiv \frac{r-r_+}{2\sqrt{M^2-a^2}}.
\end{equation}
%~~~~~~~~~~~~~~~~~~~~~~~~~~~~~~~~~~~~~~~~~~~~~
The relation between $\Delta$ and $z$ is $\Delta=4(M^2-a^2)z(z+1)$
(note that as $a\to 0$ both $\Delta$ and $z$ coincide with their above
Schwarzschild definitions).
One can now verify that for $m=0,\omega=0$ the master equation takes
exactly the form of Eq.\ (\ref{eq10z}) \cite{Kerr}.
Therefore, the two static solutions in the Kerr case are exactly
$\phi_r$ and $\phi_{ir}$ defined above (viewed as functions of $z$). We
already know that the solution $\phi_r$ (like $\Delta^s \phi_r$) is a
perfectly regular polynomial of $z$ (and hence of $r$), whereas the
solution $\phi_{ir}$ (like $\Delta^s \phi_{ir}$) includes a term
proportional to $\ln z$ and is hence irregular at the EH.

Let us summarize the above results concerning the regularity of static
(i.e.\ $w=0$) modes:\\
(i) {\em The case $m\neq 0$}: For both $s<0$ and $s>0$, the regular
solution is $\psi_b$ (just as in the $w\ne 0$ Schwarzschild case).
The field associated with this regular asymptotic solution is given in
Eq.\ (\ref{eq114a20}).
\\
(ii) {\em The case $m=0$}:
For both $s<0$ and $s>0$, the regular solution is $\phi_r$ -- just as in
the static Schwarzschild case.
The field associated with this regular solution is
%~~~~~~~~~~~~~~~~~~~~~~~~~~~~~~~~~~~~~~~~~~~~~
\begin{equation} \label{eq114a30}
Y^{s,l,m=0}(\theta)\, \phi_r (z),
\end{equation}
%~~~~~~~~~~~~~~~~~~~~~~~~~~~~~~~~~~~~~~~~~~~~~
with the function $\phi_r(z)$ given in Eq.\ (\ref{eq11}),
and its asymptotic behavior (for both positive and negative $s$)
given in Eq.\ (\ref{eq12}).
[Note that in terms of the limit $m=0$ of
Eqs.\ (\ref{eq114a10},\ref{eq114a20}), Eq.\ (\ref{eq114a30})
conforms with $\Psi_a$ for $s>0$ and with $\Psi_b$ for $s<0$.]

The above results (whose detailed derivation is given in \cite{Kerr})
are summarized in Table I.
This table displays the asymptotic form of the regular and
irregular static modes for the various possible values of $s,m$.

After we have discussed the regularity features of the static solutions,
we are in a position to analyze the decay rate of the late time tails
along the EH, using the late-time expansion. As we mentioned
above, the master equation for the Kerr background is
only separable in the frequency domain. Since the late-time expansion
is carried out in the time domain, we cannot take advantage of the full
separability of the field equation. The dependence on $\varphi$ is still
separable via $e^{im\varphi}$, however, and without loss of generality
we shall consider a field $\Psi^{sm}$ of a single $m$ (the overall
perturbation field will be obtained by a superposition of all $m$
values).
To deal with the dependence on $\theta$, we proceed as follows:
We first perform the late-time expansion of the full perturbation field
$\Psi^{sm}$, and then, for each term $k$ in this expansion, we separate
the angular
dependence by decomposing into spin-weighted {\it spherical}
harmonics. The full decomposition thus takes the form
%~~~~~~~~~~~~~~~~~~~~~~~~~~~~~~~~~~~~~~~~~~~~~
\begin{equation} \label{eq115a}
\Psi^{sm} =\sum_{k=0}^{\infty}\left[\sum_{l= l_0}^{\infty}
Y^{slm}(\theta,\varphi)\,F_k^{slm}(r)\right] v^{-k_0-k},
\end{equation}
%~~~~~~~~~~~~~~~~~~~~~~~~~~~~~~~~~~~~~~~~~~~~~
where $l_0$ is the minimal value of $l$ allowed for the mode $m,s$ in
question, that is, $l_0=max(|m|,|s|)$.
The parameter $k_0$ is defined here to be the dominant late-time power index
of $\Psi^{sm}$ along lines of constant $r>r_+$; Namely, it is determined
by the multipole $l$ which has the slowest decay at $r={\rm const}>r_+$.
Note that by this definition, $k_0$ is independent of $l$ (unlike
the Schwarzschild case, in which the late-time expansion was
implemented for each mode $l,m$ in separate).
An investigation of the late-time decay at fixed $r$ \cite{Kerr,HodII}
indicates that generically the dominant multipole is the one with the
smallest $l$ allowed, i.e. $l=l_0$, and its decay rate (at fixed $r>r_+$)
is $v^{-2l_0-3}$, with all other multipoles decaying faster. This means
that generically $k_0=2l_0+3$, and also, the term $k=0$ includes only
one multipole, $l=l_0$ (that is, $F_{k=0}^{slm}$ vanishes for all $l>l_0$).

When the expression (\ref{eq115a}) is substituted in the master
equation \cite{Teukolsky73}, one finds that the radial functions
$F_k^{slm}$ still admit equations of the form (\ref{eq5}).
However, in the Kerr case the source term $S_k^{slm}$ involves
also contributions from other values of $l$.
(Actually, the source term $S_k^{slm}$ couples a multipole $l$ to
multipoles $l\pm 1,l\pm 2$.) Still, one finds that, as
in the Schwarzschild case, $S_k^{slm}$ depends only
on functions $F_{k'}$ with $k'<k$ \cite{Kerr}. In particular, the function
$F_{k=0}^{slm}$ has no source term, so it satisfies a closed
homogeneous equation, which is just the {\it static field equation}.
This structure allows one to solve for all unknowns
$F_{k}^{slm}$ in an inductive manner, starting with the functions
$F_{k=0}^{slm}$.
(The situation here is analogous to that of a scalar field in a Kerr
spacetime, analyzed in Ref.\ \cite{Letter}.)

As we have just explained, the function $F_{k=0}$
must be a static solution of the master equation. Furthermore, the
regularity arguments discussed
in Sec.\ \ref{behavior} (for the Schwarzschild case) are applicable to
the Kerr case as well, and
imply that $F_{k=0}$ must be the {\it regular} static solution. The
decay rate of the late-time tail along the Kerr EH now follows
immediately from the above discussion of the regular static solutions,
as we now describe.

For $m\neq 0$, the regular static solution is $\psi_b$.
Since $\psi_b (r)$ has the maximal amplitude allowed by the regularity
criterion, the terms $k\geq 1$ (being proportional to
$v^{-k_0-k}$) will be negligible. Therefore, the late-time tail at the EH
will be proportional to $\psi_b (r) v^{-k_0}$, and
for both $s<0$ and $s>0$ we shall have
(to the leading order in $\Delta$ and $1/v$)
%~~~~~~~~~~~~~~~~~~~~~~~~~~~~~~~~~~~~~~~~~~~~~
\begin{equation} \label{eq116a}
\Psi^{sm}\propto Y^{s,l_0,m}(\theta,\tilde\varphi_+)\,
e^{im\Omega_+v}\,\Delta^{-s}\, [v^{-k_0}+O(v^{-k_0-1})],\;\;\;\;
\text{for $m\neq 0$}.
\end{equation}
%~~~~~~~~~~~~~~~~~~~~~~~~~~~~~~~~~~~~~~~~~~~~~
(The oscillatory factor $e^{im\Omega_+v}$ has already been observed in
the scalar-field case \cite{Amos,Letter}.)
Note that the angular dependence in this expression, as well as in Eqs.\
(\ref{eq116b},\ref{eq117}) below, only includes the multipole $l_0$: As
was mentioned above, generically $F_{k=0}^{slm}$ vanishes for all
$l>l_0$.

For $m=0$, the situation is just as in the Schwarzschild case:
In the case $s<0$, the regular static solution $\phi_r$ is proportional
to $\Delta^{-s}$. Since this is the maximal magnitude allowed by the
regularity criterion, the terms $k\geq 1$ will be negligible in this case
too (just as in the case $m\neq 0$ above). The late-time tail at the EH
will therefore be proportional to
$\phi_r (r) v^{-k_0}$, and we obtain
%~~~~~~~~~~~~~~~~~~~~~~~~~~~~~~~~~~~~~~~~~~~~~
\begin{equation} \label{eq116b}
\Psi^{sm}\propto Y^{s,l_0,m=0}(\theta)\,\Delta^{-s}\,
[v^{-k_0}+O(v^{-k_0-1})],\;\;\;\;
\text{for $m=0,s<0$}.
\end{equation}
%~~~~~~~~~~~~~~~~~~~~~~~~~~~~~~~~~~~~~~~~~~~~~
However, in the case $s>0$ (yet with $m=0$), the $k=0$ term is
proportional to the regular static solution
$\phi_r\propto \Delta^0$, whereas the $k=1$ term diverges like
$\Delta^{-s}$. Therefore, at the EH the $k=1$ term dominates, and the
late-time asymptotic behavior near the EH is given by
%~~~~~~~~~~~~~~~~~~~~~~~~~~~~~~~~~~~~~~~~~~~~~
\begin{equation} \label{eq117}
\Psi^{sm}\propto Y^{s,l_0,m=0}(\theta)\,
[v^{-k_0}+\bar c\, \Delta ^{-s}\,
v^{-k_0-1} +O(v^{-k_0-2})], \;\;\;\;
\text{for $m=0,s>0$}.
\end{equation}
%~~~~~~~~~~~~~~~~~~~~~~~~~~~~~~~~~~~~~~~~~~~~~
In Ref. \cite{Kerr} we verify (by calculating $F_{k=1}$) that the
coefficient $\bar c$ is non-vanishing, and also that the term
proportional to $\Delta ^{-s}\,v^{-k_0-1}$ includes the multipole
$l=l_0$ only.
Note that for $m=0,s>0$, the decay rate along the EH is $v^{-k_0-1}$,
whereas in all other cases it is $v^{-k_0}$. (The decay rate along lines
of constant $r>r_+$ is $v^{-k_0}$ in all cases.)

As was mentioned above, generically $k_0=2l_0+3$. Thus, the most
dominant $m$-modes are those with $|m|\leq |s|$. For these $m$-modes
$l_0=|s|$, so $k_0=2|s|+3$. Therefore, in the {\it overall} perturbation
field $\Psi^{s}$, made of the superposition of all $m$-modes, the decay
rate along the EH is generically $v^{-2|s|-3}$. Recall, however, that for
$s>0$ fields the axially-symmetric component ($m=0$) decays faster
along the EH, like $v^{-2|s|-4}$.

Note that the discussion is this section only deals with
non-extremal, $|a|<M$, Kerr BHs: the extremal case needs to be considered
separately. (Note, for example, that Eq.\ (\ref{zKerr}) is not valid in
the case $a=M$.)

\section{conclusions}\label{conclusions}

To summarize our results, it is most useful to refer to the
physical variables $\hat\Psi^s=\Delta^s \Psi^s$ defined above.
These variables are natural, because, as we
mentioned above, at the EH they are proportional to the regular Maxwell
components $F_{aV}$ or $F_{aU}$, or to the regular Weyl components
$C_{aVbV}$ or $C_{aUbU}$ (for $s=1$, $s=-1$, $s=2$, and $s=-2$,
respectively; here, $a,b$ stand for the two regular angular
coordinates, i.e. $\theta,\varphi$ in the Schwarzschild case and
$\theta,\tilde\varphi_+$ in the Kerr case).

{\em For the Schwarzschild case}, we find from Eqs.\ (\ref{eq59}) and
(\ref{eq62}) that along the
EH, each mode $l,m$ of the physical field $\hat\Psi^s$
decays at late time with the leading-order tail
%~~~~~~~~~~~~~~~~~~~~~~~~~~~~~~~~~~~~~~~~~~~~~
\begin{equation} \label{eq118}
\left.\begin{array}{lcl}
\hat\psi^{slm}(v)&\cong &{\rm const}\cdot \,v^{-2l-3+p},\quad
\text{for $s<0$}\\
\hat\psi^{slm}(v)&\cong &{\rm const}\cdot \,v^{-2l-4+p},\quad
\text{for $s>0$}
\end{array}\right\}\quad \text{Schwarzschild BH},
\end{equation}
%~~~~~~~~~~~~~~~~~~~~~~~~~~~~~~~~~~~~~~~~~~~~~
where the constants are generically non-vanishing.
Here, $p=1$ if there exists an initial static multipole
$l$, or $p=0$ in the absence of such a multipole.
(In the calculations above we assumed $p=0$,
but the extension to the $p=1$ case is trivial.)
For the comparison with the Kerr case below
(in which a full angular separability
of the late-time tails is not possible), it is useful to re-write
Eq.\ (\ref{eq118}) in terms of the field $\hat\Psi^{sm}$, which is the
part of $\hat\Psi^s$ including all multipoles $l$ for a given $m$.
$\hat\Psi^{sm}$ is dominated by the minimal multipole $l$ allowed for
the values $s,m$, i.e. $l_0\equiv max(|m|,|s|)$. Equation (\ref{eq118})
yields
%~~~~~~~~~~~~~~~~~~~~~~~~~~~~~~~~~~~~~~~~~~~~~
\begin{equation} \label{eq118m}
\left.\begin{array}{lcl}
\hat\Psi^{sm}\cong {\rm const}\cdot Y^{s,l_0,m}(\theta,\varphi) \,
v^{-2l_0-3+p},\quad
\text{for $s<0$}\\
\hat\Psi^{sm}\cong {\rm const}\cdot Y^{s,l_0,m}(\theta,\varphi) \,
v^{-2l_0-4+p},\quad
\text{for $s>0$}
\end{array}\right\}\quad \text{Schwarzschild BH}.
\end{equation}
%~~~~~~~~~~~~~~~~~~~~~~~~~~~~~~~~~~~~~~~~~~~~~
This holds for both axially-symmetric and nonaxially-symmetric
modes.

{\em For the (nonextremal) Kerr case}, we find from Eqs.\
(\ref{eq116a})---(\ref{eq117})
that the physical field $\hat\Psi^{sm}$ for a given value
of $m$ decays along the EH according to
%~~~~~~~~~~~~~~~~~~~~~~~~~~~~~~~~~~~~~~~~~~~~~
\begin{equation} \label{eq118a}
\hat\Psi^{sm}\cong {\rm const}\cdot
Y^{s,l_0,m}(\theta,\tilde\varphi_+)\,
e^{im\Omega_+v}\,v^{-2l_0-3+p},
\quad \text{Kerr BH, $m\neq 0$},
\end{equation}
%~~~~~~~~~~~~~~~~~~~~~~~~~~~~~~~~~~~~~~~~~~~~~
for nonaxially-symmetric modes, and
%~~~~~~~~~~~~~~~~~~~~~~~~~~~~~~~~~~~~~~~~~~~~~
\begin{equation} \label{eq118b}
\left.\begin{array}{lcl}
\hat\Psi^{sm}&\cong &{\rm const}\cdot Y^{s,l_0,m=0}(\theta)\,
v^{-2l_0-3+p},\quad \text{for $s<0$}\\
\hat\Psi^{sm}&\cong &{\rm const}\cdot Y^{s,l_0,m=0}(\theta)\,
v^{-2l_0-4+p},\quad \text{for $s>0$}
\end{array}\right\}\quad \text{Kerr BH, $m=0$},
\end{equation}
%~~~~~~~~~~~~~~~~~~~~~~~~~~~~~~~~~~~~~~~~~~~~~
for axially-symmetric modes.

The {\it overall} late-time behavior of the field $\hat\Psi^{s}$ is
dominated
by the $m$-values which yield the minimal possible value of $l_0$,
i.e. the values $-|s|\leq m\leq |s|$, for which $l_0=|s|$.
Thus, the overall decay rate of $\hat\Psi^{s}$ is obtained from Eqs.\
(\ref{eq118m},\ref{eq118a},\ref{eq118b}) by substituting $l_0\to |s|$.
Note, however, that the angular dependence of the overall late-time
field (and, in the Kerr case, also the oscillations in $v$) will be
obtained by a superposition over all $m$
values in the range $-|s|\leq m\leq |s|$.

The late-time behavior of an $s>0$ field along the EH of the Kerr black
hole
displays two important differences between the axially-symmetric and
nonaxially-symmetric modes. First, the modes $m\neq 0$ oscillate
along
the horizon's generators according to $e^{im\Omega_+v}$, whereas the
mode $m=0$ decays monotonically.
Second, the modes $m\neq 0$ (with $|m|\leq s$) decay like $v^{-2s-3}$,
whereas the mode $m=0$ decays along the EH like $v^{-2s-4}$.
(The first difference applies also for $s\leq 0$ fields, but the second
one
is special to $s>0$.)
These two differences lead to an interesting consequence:
The {\it overall} $s>0$ late-time field oscillates along the generators
of
the Kerr EH (because the non-oscillatory mode $m=0$ decays faster
than the oscillatory modes $m\neq 0$).
On the other hand, for $s<0$ fields the overall late-time tail at the
Kerr EH is a superposition of oscillatory $m\neq 0$ tails and a
monotonic
$m=0$ tail (which decays at the same rate).

The above difference between the $s<0$ and the $s>0$ fields
in the power-law indices of the tails along the EH has never been
reported before (as far as we know), even in the Schwarzschild case.
In Ref.\ \cite{HodII} Hod attempts to calculate these tails in the
Kerr case, but his analysis yields no difference between the $s<0$
and $s>0$ fields, even for $m=0$ (see, however, the note added).
In what follows we briefly explain what seems to be the reason for
the incorrect result in Ref. \cite{HodII}.

Hod uses the correct asymptotic radial solution
for $\omega\neq 0$ modes near the EH, which for $m=0$ reads
$\psi_{\omega}(r)\cong C(\omega)\Delta^{-s}e^{-i\omega r_*}$
(see Eq.\ (39) in \cite{HodII}
\footnote
{Note that in \cite{HodII} a different dependent field variable
is used, which is $\Delta^{s/2}$ times
the Newman-Penrose fields that we use in the present paper.}).
He then continues by assuming that $C(\omega)$ is
$\omega$-independent.
This assumption, however, is invalid in the case
$am=0$,$s>0$, where $C(\omega)\propto \omega$.
In the Schwarzschild case (for any $m$),
this result was demonstrated in the framework
of our toy-model in Sec.\ \ref{toy} [see Eq.\ (\ref{eq103})],
and can also be verified
for the actual $s>0$ fields by using the Teukolsky-Starobinsky
identities [see Eqs.\ (\ref{eq111},\ref{eq112})].
In the Kerr case (for $m=0$), the situation seems to be the
same.\footnote{
As was shown in Sec.\ \ref{Kerr}, for $m=0,s>0$ (in Kerr) the
regular solution switches from $\psi_b$ to $\psi_a$ as $\omega$
vanishes
(just like for $s>0$ in the Schwarzschild case).
This switching seems to indicate the vanishing of $C(\omega)$ as
$\omega\to 0$, as discussed in Sec.\ \ref{toy}.
[Note that for $m\neq 0,s>0$, the
regular solution is $\psi_b$ for both $\omega=0$ and $\omega\neq 0$
(we are only interested here in small $\omega$ values near
$\omega=0$,
so we can now assume $\omega -m\Omega_+\neq 0$),
indicating that in this case
$C(\omega)$ is non-vanishing as $\omega\to 0$.
The same holds for $s<0$ (and any $m$).]
}
If the correct form, $C(\omega)\propto \omega$, is used in the case
$m=0,s>0$,
then, Eq.\ (40) in \cite{HodII} correctly yields
a tail smaller by $1/v$ than the $s<0$ tail.

The asymptotic behavior of the various physical fields along the EH
is important for understanding the dynamics of these fields {\it inside}
the black hole:
One can naturally view the field value at the EH as initial data for
the black-hole's interior. (This is most naturally implemented within
the framework of the characteristic initial-value formulation.)
Of special importance is the case of gravitational perturbations
($s=\pm 2$)
of the Kerr background. In this case, evolving the
perturbation from the EH to the future, determines the
gravitational field (and hence the spacetime geometry) inside the
black hole, up to the inner horizon.
The infinite blue-shift of the gravitational perturbations
leads to a curvature singularity at the inner horizon
\cite{InnerRotating}.
Knowing the late-time behavior of the perturbation
along the Kerr EH enables one to
analyze in detail the structure of this singularity \cite{Ori99}.

{\em Note added}.---After this paper was submitted, Hod presented the
correct result for $s>0$ fields (the one derived in Sec.\ \ref{Kerr}
above), in a recent manuscript \cite{HodIII}.

\appendix

%****************************  APPENDIX A ********************************
\section{Derivation of $\Psi^{\lowercase{s}>0}$ from
$\Psi^{\lowercase{s}<0}$ using the Teukolsky-Starobinsky identities}
\label{appA}
%*************************************************************************

In this appendix we obtain the asymptotic form of $\Psi^{s>0}$
at the EH of a Schwarzschild BH from that of $\Psi^{s<0}$,
using the Starobinsky-Teukolsky identities.
By this we shall recover Eq.\ (\ref{eq62}),
and show that the leading-order coefficient $c_1^+$
does not vanish. (This was already verified in Sec.\ (\ref{behavior})
by a direct calculation of $\Psi^{s>0}$ from the field equation for
$s>0$.)

The Teukolsky-Starobinsky identities \cite{TS} relate the
perturbation fields $\Psi^{s>0}$ and $\Psi^{s<0}$.
In the case of the Schwarzschild background, these identities
take the form
%~~~~~~~~~~~~~~~~~~~~~~~~~~~~~~~~~~~~~~~~~~~~~~~~~~~~~~~~~~~~~~~~~~~~~~
\begin{equation} \label{eqA64}
(a_s -i\omega b_s\partial_t)\,\psi_{\omega}^s(r)=
{\cal D}_0^{2s}[\psi_{\omega}^{-s}(r)] \quad \quad (s>0),
\end{equation}
%~~~~~~~~~~~~~~~~~~~~~~~~~~~~~~~~~~~~~~~~~~~~~~~~~~~~~~~~~~~~~~~~~~~~~~
where $\psi_{\omega}^s(r)$ is the radial Fourier mode introduced
in Eq.\ (\ref{eq81}),  $a_s$ is a non-vanishing constant,
$b_s$ is a constant which vanishes for $s=1$ but not for $s=2$,
and ${\cal D}_0$ is a differential operator given by
%~~~~~~~~~~~~~~~~~~~~~~~~~~~~~~~~~~~~~~~~~~~~~~~~~~~~~~~~~~~~~~~~~~~~~~
\begin{equation} \label{eqA1}
{\cal D}_0(r)\equiv \partial_r -i\omega r^2\Delta^{-1}.
\end{equation}
%~~~~~~~~~~~~~~~~~~~~~~~~~~~~~~~~~~~~~~~~~~~~~~~~~~~~~~~~~~~~~~~~~~~~~~
(There also is an analogous identity for transforming from
$s>0$ to $s<0$.)
If we now apply the inverse Fourier transform to Eq.\ (\ref{eqA64}),
we obtain for the time domain function $\psi^{slm}(r,t)$
[the one introduced in Eq.\ (\ref{eq1})]
%~~~~~~~~~~~~~~~~~~~~~~~~~~~~~~~~~~~~~~~~~~~~~~~~~~~~~~~~~~~~~~~~~~~~~~
\begin{equation} \label{eqA65}
(a_s+b_s\,\partial_t)\,\psi^{slm}=
\hat{\cal D}_{s}[\psi^{-s,lm}]  \quad\quad (s>0).
\end{equation}
%~~~~~~~~~~~~~~~~~~~~~~~~~~~~~~~~~~~~~~~~~~~~~~~~~~~~~~~~~~~~~~~~~~~~~~
where $\hat{\cal D}_{s}$ is the differential operator
%~~~~~~~~~~~~~~~~~~~~~~~~~~~~~~~~~~~~~~~~~~~~~~~~~~~~~~~~~~~~~~~~~~~~~~
\begin{equation} \label{eqA65a}
\hat{\cal D}_{s}(r,t)\equiv (\partial_r +r^2 \Delta^{-1}\partial_t)^{2s}=
(2r^2\Delta^{-1}\partial_v)^{2s}.
\end{equation}
%~~~~~~~~~~~~~~~~~~~~~~~~~~~~~~~~~~~~~~~~~~~~~~~~~~~~~~~~~~~~~~~~~~~~~~
Here, the $v$-derivative is taken with fixed $u$,
and the $t$-derivative is taken with fixed $r$.

Before making use of this identity to study the late-time tails,
let us briefly discuss its application to the static
solutions. In the static case, Eq.\ (\ref{eqA65}) reduces to
%~~~~~~~~~~~~~~~~~~~~~~~~~~~~~~~~~~~~~~~~~~~~~~~~~~~~~~~~~~~~~~~~~~~~~~
\begin{equation} \label{eqA66}
a_s\psi ^{slm}=(\partial _r)^{2s}\,\psi ^{-s,lm}
\quad\quad \text{($s>0$, static)}.
\end{equation}
%~~~~~~~~~~~~~~~~~~~~~~~~~~~~~~~~~~~~~~~~~~~~~~~~~~~~~~~~~~~~~~~~~~~~~~
Consider first the application of this identity to the regular
static solution $\phi_r^-$ (namely, we take $s>0$ and $\psi
^{-s,lm}=\phi_r^-$). Since $\phi_r^-$ is a polynomial in $r$, the
right-hand side must be a polynomial too. Since the outcome must
be a static solution for $s>0$, it must be the polynomial static
solution, i.e.\ $\phi_r^+$ (up to some constant).
This confirms our previous conclusion,
namely, that for $s>$0 the regular static solution is $\phi_r^+$
and not $\phi_{ir}^+$ (i.e.\ the one proportional to $\Delta^0$
and not to $\Delta^{-s}$).

Next, consider the application of the identity (\ref{eqA66})
to the other, irregular, static solution $\phi_{ir}^-$. When
the differential operator $(\partial_r)^{2s}$ acts on the
logarithmic-free
terms in the right-hand side of Eq.\ (\ref{eq45}) (2nd row),
it yields a regular polynomial, as before. However, when applied
to the logarithmic term, it yields two types of terms:
(i) logarithmic terms, proportional to $\Delta ^0$ -- these
are obtained if the derivative operator never acts on the
factor $\ln (z)$.
(ii) Non-logarithmic terms proportional to negative (as
well as positive) powers of $\Delta$: These terms are
obtained when one of the operators $\partial_r$ acts on
the factor $\ln (z)$. The most dominant negative power
is obtained when $\partial_r$ acts on $\ln (z)$ on its
$|s|+1$ operation, which yields a contribution proportional
to $\Delta^{-s}$ (all other contribution are of less
negative powers of $\Delta$).
One can identify the terms (i) and (ii) with the second and
first terms, respectively, in the 1st row on the right-hand side
of Eq.\ (\ref{eq45}).
Note the crucial role played by the logarithmic term
in $\phi_{ir}^-$  (despite the fact that it only appears
in a sub-dominant term proportional to $\Delta ^{|s|}$):
Without this logarithmic term, the operation in Eq.\ (\ref{eqA66})
would have yielded a perfectly smooth function of $r$, proportional
to $\Delta ^0$.

We shall now apply the (time-domain) Teukolsky-starobinsky identity
(\ref{eqA65}) to the $s<0$ late-time field (\ref{eq58}).
For the consistency of the notation, we shall assume that
$s>0$ [as dictated by the notation of Eq.\ (\ref{eqA65})],
and therefore re-write Eq.\ (\ref{eq58}) as
%~~~~~~~~~~~~~~~~~~~~~~~~~~~~~~~~~~~~~~~~~~~~~~~~~~~~~~~~~~~~~~~~~~~~~~
\begin{equation} \label{eqA67}
\psi^{-s,lm}(t,r)\cong c_0^-\,\phi_r^-(r)\,v^{-2l-3}
\quad \quad (s>0).
\end{equation}
%~~~~~~~~~~~~~~~~~~~~~~~~~~~~~~~~~~~~~~~~~~~~~~~~~~~~~~~~~~~~~~~~~~~~~~
Applying the differential operator $\hat{\cal D}_s$ to the
right-hand side of Eq.\ (\ref{eqA67}), we obtain three types of
terms:\\
(i) a term in which the derivative operator $\partial _v$
never acts on $v^{-2l-3}$,\\
(ii) a term in which the derivative operator $\partial _v$
acts on $v^{-2l-3}$ once,\\
(iii) terms in which the derivative operator $\partial _v$
acts on $v^{-2l-3}$ more than once.

Consider first the term (i). As was discussed above,
the operator $\hat{\cal D}_s$ transforms $\phi_r^-$ into $\phi_r^+$.
Hence the term (i) is nothing but
%~~~~~~~~~~~~~~~~~~~~~~~~~~~~~~~~~~~~~~~~~~~~~~~~~~~~~~~~~~~~~~~~~~~~~~
\begin{equation} \label{eqA68}
c_0^-\,v^{-2l-3}\,\hat{\cal D}_s(\phi_r^-)={\rm const}\cdot v^{-2l-3}
\cdot \phi_r^+
\end{equation}
%~~~~~~~~~~~~~~~~~~~~~~~~~~~~~~~~~~~~~~~~~~~~~~~~~~~~~~~~~~~~~~~~~~~~~~
(with a non-vanishing constant).

The term (ii) is proportional to $v^{-2l-4}$. We must
keep this term, however, because its radial function will appear
to diverges at the EH [whereas the radial function in the term
(i) is regular]. The regular static solution $\phi_r^-$ is,
at the leading order, proportional to $\Delta ^s$ (recall
that now $s>0$). The most divergent contribution is obtained when
the differential operator $\partial_v$ acts on $v^{-2l-3}$ on
its $|s|+1$ application (with the other contributions smaller
by factors of $\Delta$). This contribution yields
%~~~~~~~~~~~~~~~~~~~~~~~~~~~~~~~~~~~~~~~~~~~~~~~~~~~~~~~~~~~~~~~~~~~~~~
\begin{eqnarray} \label{eqA70}
\lefteqn{\partial_v(v^{-2l-3})\,(\partial_r)^{s-1}[2r^2\Delta^{-1}
(\partial_r)^s(\Delta^s)] \cong}  \nonumber\\
&&\left[2(2l+3)(-1)^s s!(s-1)! (2M)^{2s+1}\right]
\Delta^{-s} v^{-2l-4}\,[1+O(\Delta)].
\end{eqnarray}
%~~~~~~~~~~~~~~~~~~~~~~~~~~~~~~~~~~~~~~~~~~~~~~~~~~~~~~~~~~~~~~~~~~~~~~
(Recall that when acted on a purely radial function,
$2r^2\Delta ^{-1}\partial _v=r^2\Delta ^{-1}
\partial _{r*}=\partial _r$.)

The terms (iii) decay as $v^{-2l-5}$ or faster. The
radial functions involved in these terms do not diverge
faster than $\Delta ^{-s}$ (in fact, they diverge even
slower). Therefore, these terms may be neglected. We find
that
%~~~~~~~~~~~~~~~~~~~~~~~~~~~~~~~~~~~~~~~~~~~~~~~~~~~~~~~~~~~~~~~~~~~~~~
\begin{equation} \label{eqA71}
\hat{\cal D}_s(\psi^{-s,lm})\cong (i)+(ii)=(\tilde c_0\,
v^{-2l-3}+\tilde c_1\Delta^{-s}\,v^{-2l-4})\,\,
[1+O(\Delta )+O(1/v)],
\end{equation}
%~~~~~~~~~~~~~~~~~~~~~~~~~~~~~~~~~~~~~~~~~~~~~~~~~~~~~~~~~~~~~~~~~~~~~~
where both constants $\tilde c_0,\tilde c_1$ are
nonvanishing and proportional to $c_0^-$.

So far we considered the contribution to $\hat{\cal D}_s(\psi^{-s,lm})$
from the term of $\psi^{-s,lm}$ proportional to $v^{-2l-3}$,
which is given in Eq.\ (\ref{eqA67}).
This leading-order term corresponds to the term $k=0$ in Eq.\
(\ref{eq4}). We now consider the contribution of the $k>0$ terms.
From Eq. (\ref{eq53}) we learn that for $s<0$ all functions
$F_k^{slm}$ are smooth functions of $r$, which vanish at least
like $\Delta ^{|s|}$ at the EH.
One can now easily analyze the contribution of
each term $k>0$ in the same way the dominant $k=0$ contribution
was analyzed above. Again one obtains contributions which
are analogous to the terms (i),(ii) or (iii) below, except that
these contributions are now multiplied by an extra factor
$v^{-k}$. The contributions from all $k>0$ terms of $\psi^{-s,lm}$
can therefore be neglected, and we are left with Eq.\ (\ref{eqA71}).

Once $\hat{\cal D}_s(\psi^{-s,lm})$ is known, we can
calculate $\psi^{slm}$ via Eq.\ (\ref{eqA65}). For $s=1$, the
coefficient $b_s$ vanishes, so
$\psi^{slm}=\hat{\cal D}_s(\psi^{-s,lm})/a_s$.
For $s=2$, the left-hand side of Eq.\ (\ref{eqA65})
includes the derivative operator $\partial _t$.
To extract $\psi^{slm}$ in this case, we apply the differential
operator $a_s+b_s\,\partial_t$ to the right-hand side of
Eq.\ (\ref{eq4}) (recalling $\partial_t\to \partial_v$), and solve
for $F_k^{slm}$ term by term by matching the powers of
$1/v$ to Eq.\ (\ref{eqA71}).
For $k=0$, this matching yields
%~~~~~~~~~~~~~~~~~~~~~~~~~~~~~~~~~~~~~~~~~~~~~~~~~~~~~~~~~~~~~~~~~~~~~~
\begin{equation} \label{eqA73}
a_s F_{k=0}^{slm}=\tilde c_0+O(\Delta ).
\end{equation}
%~~~~~~~~~~~~~~~~~~~~~~~~~~~~~~~~~~~~~~~~~~~~~~~~~~~~~~~~~~~~~~~~~~~~~~
For $k=1$ we obtain
%~~~~~~~~~~~~~~~~~~~~~~~~~~~~~~~~~~~~~~~~~~~~~~~~~~~~~~~~~~~~~~~~~~~~~~
\begin{equation} \label{eqA74}
a_s F_{k=1}^{slm}-(2l+3)b_s F_{k=0}^{slm}=
\tilde c_1\Delta^{-s}[1+O(\Delta )],
\end{equation}
%~~~~~~~~~~~~~~~~~~~~~~~~~~~~~~~~~~~~~~~~~~~~~~~~~~~~~~~~~~~~~~~~~~~~~~
which, in view of Eq.\ (\ref{eqA73}), we simply write as
%~~~~~~~~~~~~~~~~~~~~~~~~~~~~~~~~~~~~~~~~~~~~~~~~~~~~~~~~~~~~~~~~~~~~~~
\begin{equation} \label{eqA75}
a_sF_{k=1}^{slm}=\tilde c_1\Delta ^{-s}[1+
O(\Delta )].
\end{equation}
%~~~~~~~~~~~~~~~~~~~~~~~~~~~~~~~~~~~~~~~~~~~~~~~~~~~~~~~~~~~~~~~~~~~~~~
We thus obtain
%~~~~~~~~~~~~~~~~~~~~~~~~~~~~~~~~~~~~~~~~~~~~~~~~~~~~~~~~~~~~~~~~~~~~~~
\begin{equation} \label{eqA72}
\psi^{slm}=(c_0^+\,v^{-2l-3}+c_1^+\Delta ^{-s}\,v^{-2l-4})
\,\,[1+O(\Delta )+O(1/ v)]\quad \quad (s>0),
\end{equation}
%~~~~~~~~~~~~~~~~~~~~~~~~~~~~~~~~~~~~~~~~~~~~~~~~~~~~~~~~~~~~~~~~~~~~~~
with $c_{0,1}^+ \equiv \tilde c_{0,1}/a_s$.

Thus, relying on the late-time behavior (\ref{eq58}) for $s<0$,
and using the Teukolsky-Starobinsky identities,
we have recovered Eq.\ (\ref{eq62}) for $s>0$ --- with non-vanishing
coefficients $c_0^+$ and $c_1^+$.

% ************************** BIBLIOGRAPHY ******************************

% *********************************************************************

%----------------------------- TABLE 2 ----------------------------------
\begin{table}
\centerline{$\begin{array}{|cc|c|c|} \hline
\mbox{}                     & \mbox{}
&\hspace{2mm}\text{irregular static solution}\hspace{2mm}
&\hspace{2mm}\text{regular static solution}\hspace{2mm}\\ \hline
am=0,      &  s>0   &  \Delta^{-s}   & \Delta^0        \\ \hline
am=0,      &  s<0   &  \Delta^0      & \Delta^{-s}     \\ \hline
am\neq 0,  &  \text{$s<0$ and $s>0$}       &
\Delta^0e^{im\Omega_+u}  & \Delta^{-s}e^{im\Omega_+v}  \\ \hline
\end{array}$
}
\caption{\protect\footnotesize The asymptotic forms of the
physically-regular and physically-irregular static solutions
at the EH in the Kerr case.
Presented are the results for the axially-symmetric ($m=0$) modes of the
fields, as well as for its nonaxially-symmetric ($m\neq 0$) modes.
[The Schwarzschild case ($a=0$) can be read from this table by
referring only to the results in the first two lines (which then
apply to all $m$).]
}
\end{table}
%-------------------------------------------------------------------------


\begin{references}
\bibitem{Hod3} In Ref \cite{HodII} Hod recently analyzed the late
time behavior of spin-$s$ fields along the EH of a Kerr BH.
However, his calculation was based on an assumption which (as we
show at the end of this paper) turns out to be incorrect in the
case $s>0$. This let to an erroneous result in this case.
\bibitem{LTE} For further information about the late-time
expansion see, e.g., Refs.\ \cite{Amos,Barack99II}. (Note, however,
the different notation in Ref.\ \cite{Amos}.)
\end{references}
\end{document}